\documentclass[11pt,twoside,final]{article}

\usepackage{times}
\usepackage[top=1.0cm, bottom=2cm, left=3.4cm, right=3.0cm]{geometry}
\usepackage{amsmath}%
\usepackage{amsfonts}%
\usepackage{lineno}
\usepackage[small]{caption}
\usepackage{calc}
\usepackage{afterpage}
\usepackage{psfrag,relsize,graphicx}
\usepackage{authblk}
\usepackage[fit,hyphenate]{truncate} 

\usepackage[round, numbers, authoryear]{natbib}

\usepackage{psfrag,amsmath,amscd,amsfonts,latexsym,epsf}
\usepackage{graphicx,multirow}

\newcommand{\orth}{\bot}

\def\bbbr{{\mathbb R}}

\begin{document}

\pagenumbering{roman}


\title{\bf Invariance of visual operations at
               the level of\\ receptive fields}


\author{Tony Lindeberg}

\affil{Department of Computational Biology,\\
        School of Computer Science and Communication,\\
        KTH Royal Institute of Technology,\\
        Stockholm, Sweden}


\date{}

\maketitle
\let\oldthefootnote\thefootnote
\renewcommand{\thefootnote}{\fnsymbol{footnote}}
\footnotetext[1]{Address for correspondence: Department of Computational Biology, School of Computer
  Science and Communication, KTH Royal Institute of Technology, SE-100
  44 Stockholm, Sweden. Email: \texttt{tony@csc.kth.se}}
\let\thefootnote\oldthefootnote

\vspace{-8mm}

\begin{abstract}
\noindent
The brain is able to maintain a stable perception although the visual stimuli vary substantially on the retina due to geometric transformations and lighting variations in the environment. This paper presents a theory for achieving basic invariance properties already at the level of receptive fields.
 
Specifically, the presented framework comprises 
(i)~local {\em scaling transformations\/} caused by objects of different size and at
different distances to the observer, 
(ii)~locally {\em linearized image deformations\/} caused by variations in the viewing direction
in relation to the object, 
(iii)~locally {\em linearized relative motions\/} between the object and the observer and 
(iv)~local {\em multiplicative intensity transformations\/} caused by illumination
variations.

The receptive field model can be derived {\em by necessity\/} from symmetry properties of the environment and leads to predictions about receptive field profiles in good agreement with receptive field profiles measured by cell recordings in mammalian vision. Indeed, the receptive field profiles in the retina, LGN and V1 are close to ideal to what is motivated by the idealized requirements. 

By complementing receptive field measurements with selection
mechanisms over the parameters in the receptive field families, it is
shown how {\em true invariance\/} of receptive field responses can be obtained under scaling transformations, affine transformations and Galilean transformations. Thereby, the framework provides a mathematically well-founded and biologically plausible model for how basic invariance properties can be achieved already at the level of receptive fields and support invariant recognition of objects and events under variations in viewpoint, retinal size, object motion and illumination.

The theory can {\em explain\/} the different shapes of receptive field
profiles found in biological vision, which are tuned to different
sizes and orientations in the image domain as well as to different
image velocities in space-time, from a requirement that the visual
system should be invariant to the natural types of image
transformations that occur in its environment.

\medskip


\end{abstract}

\newpage

\section*{Author summary}

Receptive field profiles registered by cell recordings have shown 
that mammalian vision has developed receptive fields tuned to
different sizes and orientations in the image domain as well as
to different image velocities in space-time.
This article presents a theoretical model by which families of
idealized receptive field profiles can be derived mathematically
from a small set of basic assumptions that correspond to 
structural properties of the environment.
The article also presents a theory for how basic invariance properties 
to variations in scale, viewing direction and relative motion can be
obtained from the output of such receptive fields,
using complementary selection mechanisms that operate over the 
output of families of receptive fields tuned to different parameters.
Thereby, the theory shows how basic invariance properties of a visual system can be 
obtained already at the level of receptive fields, and we can explain
the different shapes of receptive field profiles found in biological
vision from a requirement that the visual system should be invariant
to the natural types of image transformations that occur in its environment.

\newpage

\pagenumbering{arabic}

\section{Introduction}

We maintain a stable perception of our environment although the
brightness patterns reaching our eyes
undergo substantial changes. This shows that our visual system
possesses invariance properties with respect to several types of image
transformations:

If you approach an object, it will change its size on the retina.
Nevertheless, the perception remains the same, 
which reflects a {\em scale invariance\/}.
It is well-known that humans and other animals have functionally important
invariance properties with respect to variations in scale.
For example, \citet{BieCoo92-ExpPhys} demonstrated that reaction times for
recognition of line drawings were independent of whether the primed
object was presented at the same or a different size as when
originally viewed.
\citet{LogPauPog95-CurrBiol} found that there are cells in the inferior
temporal cortex (IT) of monkeys for which the magnitude of the cell's
response is the same whether the stimulus subtended $1^{\circ}$ or
$6^{\circ}$ of visual angle.
\citet{ItoTamFujTan95-JNeuroPhys} found that about 20 percent of
anterior IT cells responded to ranges of size variations greater than 4 octaves, whereas
about 40 percent responded to size ranges less than 2 octaves.
\citet{FurEng00-VisRes} found that learning with application to object
recognition transfers across changes in image size.
The neural mechanisms underlying object recognition are rapid and lead to
scale-invariant properties as soon as 100--300 ms after stimulus onset
\citep{HunKrePogDiC05-Science}.

In a similar manner, if you rotate an object in front of you, the projected brightness
pattern will be deformed on the retina, typically by different amounts
in different directions.
To first order of approximation, such image deformations can be modelled by 
local {\em affine transformations\/}, which include the effects of
in-plane {\em rotations\/} and perspective {\em foreshortening\/}.
For example, \citet{LogPauPog95-CurrBiol} and
\citet{BooRol98-CerebrCort} have shown that in the monkey IT
cortex there are both neurons that respond selectively to particular
views of familiar objects as well as populations
of single neurons that have view-invariant
representations over different views of familiar objects.
\citet{EdeBul92-VisRes} have on the other hand shown that the time
for recognizing unfamiliar objects from novel views increases with the
3-D rotation angle between the presented and previously seen views.
Still, subjects are able to recognize unfamiliar objects from novel
views, provided that the 3-D rotation is moderate.

If an object moves in front of you, it may in addition to a
{\em translation\/} also lead to a time-dependent motion field in
the brightness pattern on the retina.
You may or you may not fixate on the object.
Depending on the relative motion between the object and the observer,
this motion field can 
be modelled by local {\em Galilean transformations\/}.
Regarding biological counterparts of such relative motions, 
\citet{RodAlb89-ExpBrainRes} and \citet{LagRaiOrb93-JNeurPhys} have shown that
in area MT of monkeys there are neurons with
high selectivity to the speed and direction of visual motion over
large ranges of image velocities.
\citet{PetBakAll85-BrainRes} have shown that there are neurons in
area MT that adapt their response properties to the direction and
velocity of motion.
\citet{SmeBre94-VisRes} have shown that reaction times for motion
perception can be different for absolute and relative motion and that
reaction times may specifically
depend on the relative motion between the object and the background.
When Einstein derived his relativity theory, he used as a basic
assumption the requirement that the equations should be invariant
under Galilean transformations \citep{Ein20-book}.

The measured luminosity of surface patterns in the world may in turn vary over
several orders of magnitude. Nevertheless we are able to preserve the
identity of an object as we move it out of or into a shade, which
reflects important invariance properties under {\em intensity transformations\/}.
The Weber-Fechner law states that the ratio of an increment
threshold $\Delta I$ in image luminosity for a just noticeable difference in relation to the background
intensity $I$ is constant over large ranges of luminosity variations
\citep[pages~671--672]{Pal99-Book}.
The pupil of the eye and the sensitivity of the photoreceptors are
continuously adapting to ambient illumination \citep{Hur02-GenPhys}.

To be able to function robustly in a complex natural world, the visual
system must be able to deal with these image transformations in an
efficient and appropriate manner to maintain a stable perception as the
brightness pattern changes on the retina.
One specific approach is by computing {\em invariant features\/} whose
values or representations remain unchanged under basic image
transformations.
A weaker but nevertheless highly useful approach is by computing
visual representations that possess suitable 
{\em covariance  properties\/}, which means that the representations
are transformed in a well-behaved and well-understood manner under
corresponding image transformations.
A covariant image representation can then in turn constitute the basis
for computing truly invariant image representations, 
and thus enable invariant visual recognition processes at the systems
level, in analogy with corresponding invariance principles as postulated for biological
vision systems by different authors \citep{Rol94-BehavProc,DiCMau00-Nature,GriRao05-NeurComp,QuiRedKreKocFri05-Nature,DiCCox07-TICS,GorBee09-FCNS}.

The subject of this paper is to introduce a computational framework
for modelling receptive fields at the earliest stages in the visual system corresponding to the
retina, LGN and V1 and to show how this framework allows for 
{\em basic invariance or covariance properties of visual operations\/} with respect to all the above mentioned phenomena.
This framework can be derived from {\em symmetry properties\/} of the natural
environment \citep{Lin10-JMIV,Lin11-RecFields} and leads to predictions of {\em receptive field profiles\/}
in good agreement with receptive measurements reported in the
literature
\citep{HubWie59-Phys,HubWie62-Phys,DeAngOhzFre95-TINS,deAngAnz04-VisNeuroSci,HubWie05-book}. 
Specifically, explicit phenomenological models will be given of LGN
neurons and simple cells in V1 and be compared to related models in terms
of Gabor functions \citep{Mar80-JOSA,JonPal87a,JonPal87b},
differences of Gaussians \citep{Rod65-VisRes} or
Gaussian derivatives \citep{KoeDoo87-BC,You87-SV,YouLesMey01-SV,YouLes01-SV}.
Notably, the evolution properties of the receptive field profiles in
this model can be described by diffusion equations and are therefore
suitable for implementation on a biological architecture,
since the computations can be expressed in terms of communications
between neighbouring computational units, where either a single
computational unit or a group of
computational units may be interpreted as corresponding to a neuron.
Specifically, computational models involving diffusion equations arise in mean field theory for
approximating the computations that are performed by populations of neurons
\cite{OmuKniSir00-CompNeuro,MatGui02-PhysRevE,FauTouCes09-FrontCompNeuroSci}.

Combined with complementary {\em selection mechanisms\/} over receptive fields
at different scales \citep{Lin97-IJCV}, receptive fields adapted to
different affine image deformations \citep{LG96-IVC} and different Galilean motions
\citep{LinAkbLap04-ICPR,Lin10-JMIV}, it will also be shown how true
invariance of receptive field responses can be obtained with respect to local scaling
transformations, affine transformations and Galilean transformations.
These selection mechanisms are based on either (i)~the computation of local
extrema over the parameters of the receptive fields or alternatively
based on (ii)~the comparisons of local receptive field responses to
affine invariant or Galilean fixed-point requirements (to be
described later).
On a neural architecture, these geometric invariance properties are
therefore compatible with a routing mechanism
\citep{OlsAndEss93-JNeurSci}
that operates on the output from families of receptive fields that are 
tuned to different scales, spatial orientations and image velocities.
In this respect, the resulting approach will bear similarity to the
approach by \citet{RiePog99-Nature}, where receptive field
responses at different scales are routed forward by a soft winner-take-all mechanism,
with the theoretical additions that the invariance
properties over scale can here be formally proven and the presented framework
specifically states
how the receptive fields should be normalized over scale.
Furthermore, our approach extends to true and provable invariance properties under
more general affine and Galilean transformations.

A direct consequence of these invariance properties established for
receptive field responses is that they can be {\em propagated to invariance
properties of visual operations at higher levels\/}, and thus enable
invariant recognition of visual objects and events under variations in
viewing direction, retinal size, object motion and illumination.
In this way, the presented framework provides a computational theory for how 
basic invariance properties of a visual system can achieved
already at the level of receptive fields. Another consequence is that
the presented framework could be used for {\em explaining\/} the families of receptive field profiles tuned to different
orientations and image velocities in space and space-time that have
been observed in biological vision from a requirement of that the
corresponding receptive field responses should be 
invariant or covariant under corresponding image transformations.
A main purpose of this article is to provide a {\em synthesis\/} where
such structural components are combined into a coherent framework for achieving basic
invariance properties of a visual system and relating these results,
which have been derived mathematically, to corresponding functional properties of
neurons in a biological vision system.


Another major aim of this article is to try to bridge the gap between
computer vision and biological vision, by demonstrating how concepts
originally developed for purposes in computer vision can be related to 
corresponding notions in computational neuroscience and biological vision.
In particular, we will argue for explicit incorporation of
basic image transformations into computational neuroscience models
of vision. If such image transformations are not appropriately modelled and
if the model is then exposed to test data that contain image variations 
outside the domain of variabilities that are spanned by the training data, 
then an artificial neuron model may have severe problems with robustness.
If on the other hand the covariance properties corresponding to the
natural variabilities in the world underlying the formation of natural image statistics are
explicitly modelled and if corresponding invariance properties are built into
the computational neuroscience model and also used in the learning stage, we argue that
it should be possible to increase the robustness of a neuro-inspired
artificial vision system to natural image variations.
Specifically, we will present explicit computational mechanism for obtaining
true scale invariance, affine invariance, Galilean invariance and
illumination invariance for image measurements in terms of local
receptive field responses.

Interestingly, the proposed framework for receptive fields can be derived 
{\em by  necessity\/} from a mathematical analysis 
based on symmetry requirements with respect to the above mentioned image
transformations in combination with a few additional requirements
concerning the internal structure and computations in the first
stages of a vision system that will be described in more detail below.
In these respects, the framework can be regarded as both 
(i)~a canonical mathematical model for the first stages of processing in an idealized
vision system and as 
(ii)~a plausible computational model for biological vision.
Specifically, compared to previous approaches of learning receptive field properties 
and visual models from the statistics of natural image data 
\citep{Fie87-JOSA,SchHat96-VisRes,OlsFie96-Nature,RaoBal98-CompNeurSyst,SimOls01-AnnRevNeurSci,Wil08-AnnRevPsychol}
the proposed theoretical model 
makes it possible to
determine spatial and spatio-temporal receptive fields from first principles 
that reflect symmetry properties of the environment 
and thus
without need for any explicit training stage or gathering of
representative image data.
In relation to such learning based models, the proposed normative
approach can be seen as describing the solutions that an ideal learning based system
may converge to, if exposed to a sufficiently large and representative
set of natural image data.
The framework for achieving true invariance properties of receptive
field responses is also theoretically strong in the sense that the
invariance properties can be formally proven given the idealized model
of receptive fields.

In their survey of our knowledge of the early visual system,
\cite{Caretal05-JNeuroSci} emphasize the need for
functional models to establish a link between neural biology and perception.
More recently,
\cite{EinKon10-CurrOpinNeuroBio} argue for the need for normative
approaches in vision.
This paper can be seen as developing the consequences of such ways of
reasoning by showing how basic invariance properties of visual
processes at the systems level can be obtained already at the 
level of receptive fields, using a normative approach.


\section{Model for early visual pathway in an idealized vision system}
\label{sec-ideal-rec-fields}

In the following we will state a number of basic requirements
concerning the earliest levels of processing in an 
idealized vision system, 
which will be used for deriving {\em idealized functional models of
receptive fields\/}.
Let us stress that the aim is not to model specific properties of 
human vision or any other species. 
Instead the goal is to describe basic characteristics of the image
formation process and the computations that are performed after the
registration of image luminosity on the retina.
These assumptions will then be used for narrowing down the class of
possible image operations that are compatible with structural requirements, 
which reflect symmetry properties of the environment.
Thereafter, it will be shown how this approach applies to modelling of
biological receptive fields and how the resulting receptive fields 
can be regarded as biologically plausible.

For simplicity, we will assume that the image measurements are performed on
a planar retina under perspective projection. With appropriate
modifications, a corresponding treatment can be performed with a
spherical camera geometry.

Let us therefore assume that the vision system receives image data
that are
either defined on a 
(i)~{\em purely spatial domain\/} $f(x)$ or a 
(ii)~{\em spatio-temporal domain\/} $f(x, t)$
with $x = (x_1, x_2)^T$.
Let us regard the purpose of the earliest levels of visual
representations as computing a family of
{\em internal representations\/} $L$ from $f$,
whose output can be used as input to different types of visual
modules.
In biological terms, this would correspond to a similar type of
{\em sharing\/} as V1 produces output for several downstream areas 
such as V2, V4 and V5/MT.

An important requirement on these early levels of processing is that
we would like them to be {\em uncommitted\/} operations without being
too specifically adapted to a particular task that would limit the
applicability for other visual tasks.
We would also desire a {\em uniform structure\/} on the first stages
of visual computations.

\subsection{Spatial (time-independent) image data}

Concerning terminology, we will use the convention that a receptive
field refers to a region $\Omega$ in visual space over which some 
computations are being performed. 
These computations will be represented by an operator ${\cal T}$,
whose support region is $\Omega$. 
Generally, the notion of a receptive field will used to refer to both the 
operator ${\cal T}$ and its support region $\Omega$.
In some cases when referring specifically to the support region
$\Omega$ only,
we will refer to it as the support region of the receptive field.

Given a purely spatial image $f \colon \bbbr^2 \rightarrow \bbbr$, 
let us consider the problem of defining a family of internal representations
\begin{equation}
  L(\cdot;\; s) = {\cal T}_s f
\end{equation}
for some family of operators ${\cal T}_s$ that are indexed by some
parameter $s$, where $s = (s_1, \cdots, s_N)$ may be a
multi-dimensional parameter with $N$ dimensions.
(The dot ``$\cdot$'' at the position of the first argument $x$ of $L$
means that that $L(\cdot;\; s)$ when given a fixed value of the parameter $s$
only should be regarded as a function over $x$.)
In the following we shall state a number of structural requirements on
a visual front-end as motivated by the types of computations that are
to be performed at the earliest levels of processing in combination
with symmetry properties of the surrounding world.

\paragraph{Linearity.}

Initially, it is natural to require ${\cal T}_s$ to be a 
{\em linear\/} operator, such that
\begin{equation}
   {\cal T}_s(a_1 f_1 + a_2 f_2) = a_1 {\cal T}_s f_1 + a_2 {\cal T}_s f_2
\end{equation}
holds for all functions $f_1, f_2 \colon \bbbr^2 \rightarrow \bbbr$ 
and all scalar constants $a_1, a_2 \in \bbbr$.
An underlying motivation to this linearity requirement is that the
earliest levels of visual processing should make as few irreversible
decisions as possible.

Linearity also implies that a number of special properties of
receptive fields (to be described below) will transfer to 
spatial and spatio-temporal derivatives of these and do therefore
imply that different types of image structures will be treated in a
similar manner irrespective of what types of linear filters they are
captured by.

\paragraph{Translational invariance.}

Let us also require ${\cal T}_s$ to be a 
{\em shift-invariant operator\/} in the sense that it commutes with
the shift operator ${\cal S}_{\Delta x}$ defined by
$({\cal S}_{\Delta x} f)(x) = f(x-\Delta x)$,
such that
\begin{equation}
   {\cal T}_s \left( {\cal S}_{\Delta x} f \right) 
   = {\cal S}_{\Delta x} \left( {\cal T}_s f \right)
  \end{equation}
holds for all $\Delta x$.
The motivation behind this assumption is the basic requirement
that the perception of a visual object should be the same 
irrespective of its position in the image plane.
Alternatively stated, the operator ${\cal T}_s$ can be said to be
{\em homogeneous across space\/}.%
\footnote{For us humans and other higher mammals, the retina is obviously
  not translationally invariant. Instead, finer scale receptive fields
  are concentrated to the fovea in such a way that the minimum receptive
  field size increases essentially linearily with eccentricity. With
  respect to such a sensor space, the assumption about translational
  invariance should be taken as an idealized model for the region in
  space where there are receptive fields above a certain size.}

\paragraph{Convolution structure.}

Together, the assumptions of linearity and shift-invariance imply that
the internal representations $L(\cdot;\; s)$ are given by {\em convolution transformations\/}
\begin{equation}
   L(x;\; s)  = (T(\cdot;\; s) * f)(x) = \int_{\xi \in \bbbr^2} T(\xi;\; s) \, f(x - \xi) \, d\xi
\end{equation}
where $T(\cdot;\; s)$ denotes some family of convolution kernels.
Later, we will refer to these convolution kernels as receptive fields.

\paragraph{The issue of scale.}

A fundamental property of the convolution operation is that it may
reflect different types of image structures depending on the spatial
extent (the width)
of the convolution kernel.
\begin{itemize}
\item
Convolution with a {\em large support\/} kernel will have the ability
to respond to phenomena at {\em coarse scales.}
\item
A kernel with {\em small support\/} may on the other hand only capture phenomena at {\em fine scales\/}.
\end{itemize}
From this viewpoint it is natural to associate an interpretation of
{\em scale\/} with the parameter $s$ and we will assume that the limit case of the internal
representations when $s$ tend to zero should correspond to original
image pattern $f$
\begin{equation}
   \label{eq-init-cond-scsp-zero-order}
    \lim_{s \downarrow 0} L(\cdot;\; s) = \lim_{s \downarrow 0} {\cal T}_s f = f.
\end{equation}

\paragraph{Semi-group structure.}

From the interpretation of $s$ as a scale parameter, it is 
natural to require the image operators ${\cal T}_s$ to form a
{\em semi-group\/}%
\footnote{Concerning the parameterization of this semi-group, we will
  in the specific case of a one-dimensional (scalar) scale parameter
  assume the parameter $s \in \bbbr$ to have a direct interpretation of scale,
whereas in the case of a multi-dimensional parameter 
$s = (s_1, \dots, s_N) \in \bbbr^N$, these parameters could also encode for
other properties of the convolution kernels in terms of the orientation
$\theta$ in image space or the degree of elongation $e = \sigma_1/\sigma_2$,
where $\sigma_1$ and $\sigma_2$ denote the spatial extents in different directions.
The convolution kernels will, however, not be be required to form a semi-group
over any type of parameterization, such as the parameters $\theta$ or
$e$.
Instead, we will assume that there exists some parameterization $s$ for
which an additive linear semi-group structure can be defined and
from which the latter types of parameters can then be derived.}
over $s$ 
\begin{equation}
  \label{eq-semi-group}
    {\cal T}_{s_1}  {\cal T}_{s_1} =  {\cal T}_{s_1 + s_2} 
\end{equation}
with a corresponding semi-group structure for the convolution kernels
\begin{equation}
    T(\cdot;\; s_1) * T(\cdot;\; s_2) = T(\cdot;\; s_1+s_2).
\end{equation}
Then, the transformation between any different and ordered scale
levels $s_s$ and $s_2$ with $s_2 \geq s_1$ will obey the 
{\em cascade property\/}
\begin{equation}
  L(\cdot;\; s_2) 
  = T(\cdot;\; s_2-s_1) * T(\cdot;\; s_1) * f 
  = T(\cdot;\; s_2-s_1) * L(\cdot;\; s_1)
\end{equation}
{\em i.e.\/} a similar type of transformation as from the original
data $f$.
An image representation with these properties is referred to as a
{\em multi-scale representation\/}.

\paragraph{Self-similarity over scale.}

Regarding the family of convolution kernels used for computing a
multi-scale representation, it is also natural to require them to 
{\em self-similar over scale\/}, 
such that if $s$ is a one-dimensional
scale parameter then all the kernels correspond to
rescaled copies
\begin{equation}
  \label{eq-self-sim-req}
   T(x;\; s) 
  = \frac{1}{\varphi(s)} \bar{T} \left(\frac{x}{\varphi(s)} \right)
\end{equation}
of some prototype kernel $\bar{T}$ for 
some transformation%
\footnote{The reason for introducing a function $\varphi$ for
  transforming the scale parameter $s$ into a scaling factor
  $\varphi(s)$ in image space, is that the requirement of a
  semi-group structure (\ref{eq-semi-group}) does not imply any restriction on how 
  parameter $s$ should be related to image measurements in dimensions
  of length --- the semi-group structure only implies an abstract
  ordering relation between coarser and finer scales $s_2 > s_1$ that
  could also be satisfied for any monotonously increasing
  transformation of the parameter $s$. 
  For the Gaussian scale-space concept with a scalar scale parameter
  and given by (\ref{eq-2D-gauss}) this transformation is given by $\sigma = \varphi(s) = \sqrt{s}$,
  whereas for the affine Gaussian scale-space concept given by
  (\ref{eq-aff-gauss-kernel}) it is given by
  the matrix square root function $\varphi(s) = \Sigma^{1/2}$, where
  $\Sigma$ denotes the covariance matrix that describes the spatial
  extent and the orientation of the affine Gaussian kernels.}
 $\varphi(s)$ of the scale parameter.
If $s \in \bbbr_+^N$ is a multi-dimensional scale parameter,
the requirement of self-similarity over scale can be generalized into
\begin{equation}
 T(x;\; s) 
 = \frac{1}{|\det \varphi(s)|} \bar{T} \left( \varphi(s)^{-1} x \right)
\end{equation}
where $\varphi(s)$ now denotes a non-singular $2 \times 2$-dimensional
matrix regarding a 2-D image domain and $\varphi(s)^{-1}$ its inverse.
With this definition, a multi-scale representation with a scalar scale 
parameter $s \in \bbbr_+$ will be based on uniform rescalings of the
prototype kernel, whereas a multi-scale representation based on a
multi-dimensional scale parameter might also allow for rotations as
well as non-uniform affine deformations of the prototype kernel.

\paragraph{Infinitesimal generator.}

For theoretical analysis it is preferable if the scale parameter can
be treated as a {\em continuous scale parameter\/} and if image
representations between adjacent levels of scale can be related by
partial differential equations. Such relations can be expressed if the
semi-group possesses an {\em infinitesimal generator\/} \citep{HilPhi57}
\begin{equation}
        {\cal B} L = \lim_{h \downarrow 0} \frac{T(\cdot;\; h) * f - f}{h}
\end{equation}
and implies that image representations between adjacent levels of
scale can be related by {\em differential evolution equations\/}; for
a scalar scale parameter of the form
\begin{equation}
        \partial_s L(x;\; s) =  ({\cal B} L)(x;\; s)
\end{equation}
for some operator ${\cal B}$ and for an $N$-dimensional scale
parameter of the form
\begin{equation}
  \label{eq-dir-der-semi-group}
  ({\cal D}_u L)(x;\; s) 
  =  ({\cal B}(u) \, L)(x;\; s) 
  = \left( u_1 {\cal B}_1 + \dots + u_N {\cal B}_N \right)
       \, L(x;\; s)
\end{equation}
for any positive direction $u = (u_1, \dots, u_N)$ in the parameter
space with $u_i \geq 0$ for every $i$.
In \citep{Lin10-JMIV} it is shown how such differential relationships can be
ensured given a proper selection of functional spaces and sufficient regularity requirements over space $x$ and
scale $s$ in terms of Sobolev norms.
We shall therefore henceforth regard the internal representations
$L(\cdot;\; s)$ as differentiable with respect to both the image space
and scale parameter(s).

\paragraph{Non-enhancement of local extrema.}

For the internal representations $L(\cdot;\, s)$ that are computed
from the original image data $f$ it is in addition essential that
operators ${\cal T}_s$ {\em do not generate new structures\/} in the
representations at coarser scales that do not correspond to simplifications of
corresponding image structures in the original image data.

\begin{figure}[hbt]
  \begin{center}
      \includegraphics[width=0.50\textwidth]{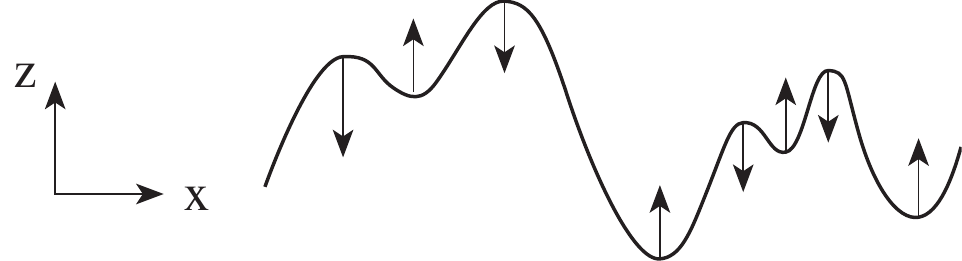}
  \end{center}
  \caption{The requirement of non-enhancement of local extrema is a
    way of restricting the class of possible image operations by
    formalizing the notion that new image structures must not
    be created with increasing scale, by requiring that the value at a
    local maximum must not increase and that the value at a local
    minimum must not decrease.}
  \label{fig-non-enh-extr}
\end{figure}

A particularly useful way of formalizing this requirement is that
{\em local extrema must not be enhanced with increasing scale\/}.
In other worlds, if a point $(x_0;\; s_0)$ is a local (spatial) maximum of the mapping
$x \mapsto L(x;\; s_0)$ then the value must not increase with scale.
Similarly, if a point $(x_0;\; s_0)$ is a local (spatial) minimum of the mapping
$x \mapsto L(x;\; s_0)$, then the value must not decrease with scale.
Given the above mentioned differentiability property with respect to
scale, we say that the multi-scale representation constitutes a
{\em scale-space representation\/} if it for a scalar scale parameter
satisfies the following conditions:
 \begin{align}
    \label{eq-non-enh-loc-extr}
    \begin{split}
      \partial_s L(x_0;\; s_0) \leq 0 \quad\quad \mbox{at any non-degenerate local maximum},
    \end{split}\\
    \begin{split}
      \partial_s L(x_0;\; s_0) \geq 0 \quad\quad \mbox{at any non-degenerate local minimum},
    \end{split}
  \end{align}
or for a multi-parameter scale-space
 \begin{align}
   \begin{split}
      ({\cal D}_u L)(x_0;\; s_0) \leq 0 \quad\quad \mbox{at any non-degenerate local maximum},
    \end{split}\\
    \begin{split}
      ({\cal D}_u L)(x_0;\; s_0) \geq 0 \quad\quad \mbox{at any non-degenerate local minimum},
    \end{split}
\end{align}
for any positive direction $u = (u_1, \dots, u_N)$ in the parameter
space with $u_i \geq 0$ for every $i$ (see figure~\ref{fig-non-enh-extr}).

\paragraph{Rotational invariance.}

If we restrict ourselves to a scale-space representation based on a scalar
(one-dimensional) scale parameter $s \in \bbbr_+$, 
then it is natural to require the scale-space
kernels to be {\em rotationally symmetric\/}
\begin{equation}
  T(x;\; s) = h(\sqrt{x_1^2 + x_2^2};\; s)
\end{equation}
for some one-dimensional function 
$h(\cdot;\; s) \colon \bbbr \rightarrow \bbbr$.
Such a symmetry requirement can be motivated by the requirement
that in the absence of further information, all spatial directions
should be equally treated (isotropy).


For a scale-space representation based on a multi-dimensional scale
parameter, one may also consider a weaker requirement of rotational invariance 
at the level of a family of kernels, for example regarding a set of 
elongated kernels with different orientations in image space.
Then, the family of kernels may capture image data of different
orientation in a rotationally invariant manner, for example if all
image orientations are explicitly represented or if the receptive fields
corresponding to different orientations in image space can be
related by linear combinations.

\paragraph{Affine covariance.}

The perspective mapping from the 3-D world to
the 2-D image space gives rises to image deformations in the image domain.
If we approximate the non-linear perspective mapping from a surface
pattern in the world to the image plane by a local linear transformation (the
derivative), then we can model this deformation by an {\em affine transformation\/}
\begin{equation}
  \label{eq-aff-def-model}
   f' = {\cal A} \, f
 \quad \mbox{corresponding to} \quad
  f'(x') = f(x) 
   \quad \mbox{with} \quad
   x' = A \, x + b.
\end{equation}
To ensure that the internal representations behave nicely under image
deformations, it is natural to require a possibility of relating them
under affine transformations
\begin{equation}
   L'(x';\; s') = L(x;\; s) 
   \quad \mbox{corresponding to} \quad
   {\cal T}_{A(s)} \, {\cal A} \, f = {\cal A} \, {\cal T}_s \, f
\end{equation}
for some transformation $s' = A(s)$ of the scale parameter.
Unfortunately, it turns out that affine covariance cannot be achieved with a scalar scale
parameter and linear operations. 
As will be shown below, it can, however, be achieved with a
3-parameter linear scale-space.

\subsection{Necessity result concerning spatial receptive fields}

Given the above mentioned requirements it can be shown that if we
assume
(i)~linearity,
(ii)~shift-invariance over space,
(iii)~semi-group property over scale,
(iv)~sufficient regularity properties over space and scale
and (v)~non-enhancement of local extrema,
then the scale-space representation over a 2-D spatial domain must
satisfy \citep[theorem 5, page 42]{Lin10-JMIV}
\begin{equation}
  \label{eq-nec-diff-eq-spat-rec-fields}
    \partial_s L 
    = \frac{1}{2} \nabla_x^T \left( \Sigma_0 \nabla_x L \right) - \delta_0^T \nabla_x L
\end{equation}
for some $2 \times 2$ covariance matrix $\Sigma_0$ 
and some 2-D vector $\delta_0$
with $\nabla_x = (\partial_{x_1}, \partial_{x_2})^T$.
If we in addition require the convolution kernels to be 
{\em mirror symmetric\/} through the origin 
$T(-x;\, s) = T(x;\; s)$ then the offset vector $\delta_0$ must
be zero.
There are two special cases within this class of operations that are
particularly worth emphasizing.

\paragraph{Gaussian receptive fields.}

\begin{figure}[hbt]
  \begin{center}
    \begin{tabular}{c}
      \includegraphics[width=0.12\textwidth]{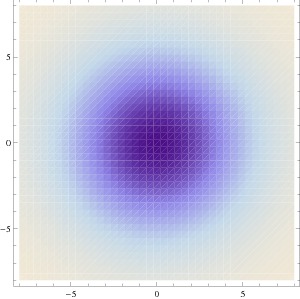} \hspace{-4mm} \\
    \end{tabular} 
  \end{center}
  \vspace{-9mm}
  \begin{center}
    \begin{tabular}{cc}
      \includegraphics[width=0.12\textwidth]{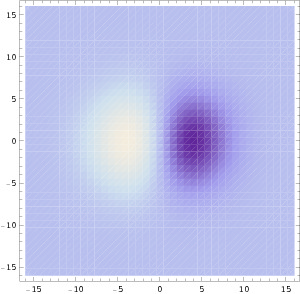} \hspace{-4mm} &
      \includegraphics[width=0.12\textwidth]{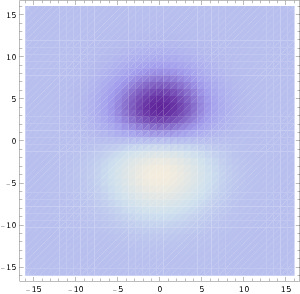} \hspace{-4mm} \\
    \end{tabular} 
  \end{center}
  \vspace{-9mm}
  \begin{center}
    \begin{tabular}{ccc}
      \includegraphics[width=0.12\textwidth]{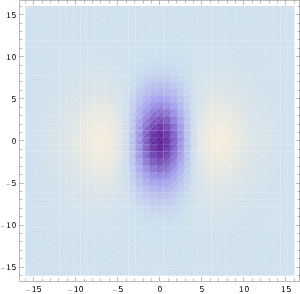} \hspace{-4mm} &
      \includegraphics[width=0.12\textwidth]{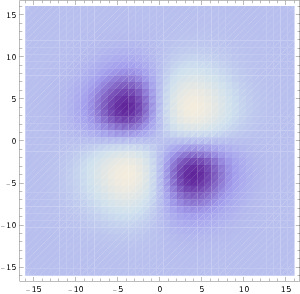} \hspace{-4mm} &
      \includegraphics[width=0.12\textwidth]{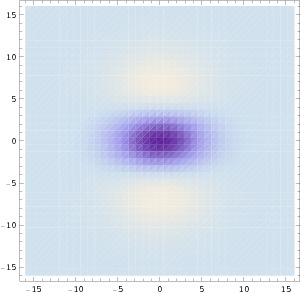} \hspace{-4mm} \\
    \end{tabular} 
  \end{center}
  \vspace{-3mm}
  \caption{Spatial receptive fields formed by the 2-D Gaussian kernel with its partial derivatives up to
    order two. The corresponding family of receptive fields is closed under
    translations, rotations and scaling transformations.}
  \label{fig-Gauss-ders-2D}
\end{figure}

If we require the corresponding convolution kernels to be rotationally
symmetric, then it follows that they will be Gaussians
\begin{equation}
  \label{eq-2D-gauss}
   T(x;\, s) = g(x;\; s) = \frac{1}{2 \pi s} \, e^{-x^T x/2 s} = \frac{1}{2 \pi s} \, e^{-(x_1^2 + x_2^2)/2 s}
\end{equation}
with corresponding {\em Gaussian derivative operators\/}
\begin{equation}
  \label{eq-gauss-ders-2D}
   (\partial_{x^{\alpha}} g)(x;\; s) 
  =  (\partial_{x_1^{\alpha_1} x_2^{\alpha_2}} g)(x_1, x_2;\; s) 
  = (\partial_{x_1^{\alpha_1}} \bar{g})(x_1;\; s) \, (\partial_{x_2^{\alpha_2}} \bar{g})(x_2;\; s)
\end{equation}
(with $\alpha = (\alpha_1, \alpha_2)$ where $\alpha_1$ and $\alpha_2$ denote the order of differentiation
in the $x_1$- and $x_2$-directions, respectively) as shown in figure~\ref{fig-Gauss-ders-2D} with the corresponding
one-dimensional Gaussian kernel and its Gaussian derivatives of the form
\begin{align}
  \begin{split}
     \bar{g}(x_1;\; s) & = \frac{1}{\sqrt{2 \pi s}} e^{-x_1^2/2s},
  \end{split}\\
  \begin{split}
    \bar{g}_{x_1}(x_1;\; s) & = - \frac{x_1}{s} \bar{g}(x_1;\; s) = - \frac{x_1}{\sqrt{2 \pi} s^{3/2}} e^{-x_1^2/2s},
  \end{split}\\
  \begin{split}
    \bar{g}_{x_1x_1}(x_1;\; s) & = \frac{(x_1^2 - s)}{s^2} \bar{g}(x_1;\; s) = \frac{(x_1^2 - s)}{\sqrt{2 \pi} s^{5/2}} e^{-x_1^2/2s}.
  \end{split}
\end{align}
Such Gaussian functions have been previously used for modelling
biological vision by \citep{You87-SV}, who has shown that there are
receptive fields in the striate cortex that can be well modelled by
Gaussian derivatives up to order four. More generally, these
Gaussian derivative operators can be used as a 
{\em general basis\/} for expressing image operations 
  such as feature detection, feature classification, surface shape,
  image matching and image-based recognition
\citep{Wit83,Koe84,KoeDoo92-PAMI,Lin93-Dis,Lin94-SI,Flo97-book,Haa04-book,Lin08-EncCompSci};
see specifically
\citep{SchCro00-IJCV,LinLin04-ICPR,Low04-IJCV,BayEssTuyGoo08-CVIU,LinLin12-CVIU}
for explicit approaches for object recognition based on Gaussian
receptive fields or approximations thereof.

\paragraph{Affine-adapted Gaussian receptive fields.}

\begin{figure}[hbt]
  \begin{center}
    \begin{tabular}{ccc}
     \includegraphics[width=0.12\textwidth]{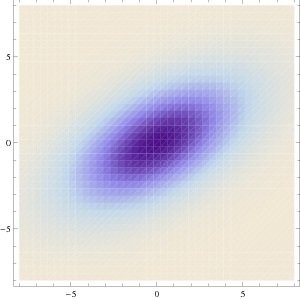} \hspace{-4mm} &
      \includegraphics[width=0.12\textwidth]{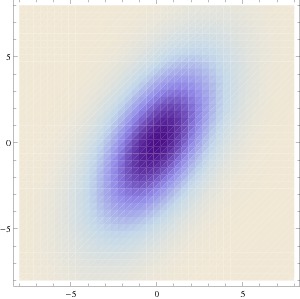} \hspace{-4mm} &
      \includegraphics[width=0.12\textwidth]{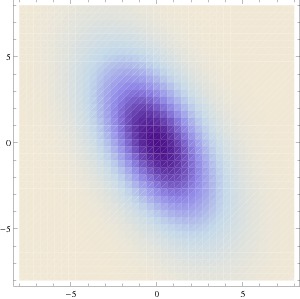} \hspace{-4mm} \\
   \end{tabular} 
  \end{center}
  \vspace{-9mm}
   \begin{center}
    \begin{tabular}{ccc}
      \includegraphics[width=0.12\textwidth]{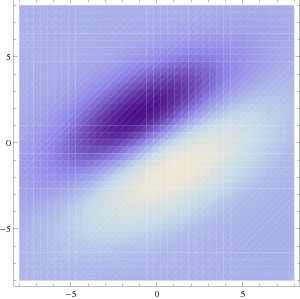} \hspace{-4mm} &
      \includegraphics[width=0.12\textwidth]{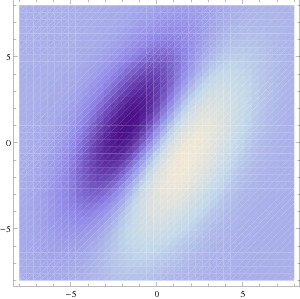} \hspace{-4mm} &
      \includegraphics[width=0.12\textwidth]{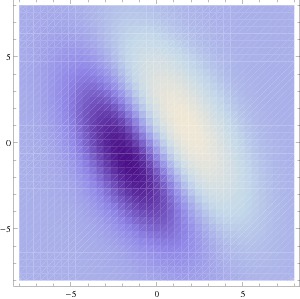} \hspace{-4mm} \\
    \end{tabular} 
  \end{center}
  \vspace{-9mm}
  \begin{center}
    \begin{tabular}{ccc}
      \includegraphics[width=0.12\textwidth]{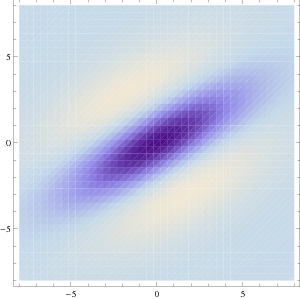} \hspace{-4mm} &
      \includegraphics[width=0.12\textwidth]{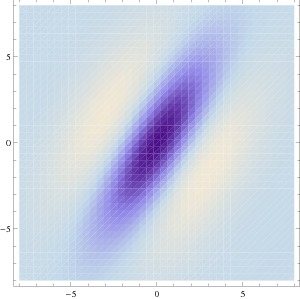} \hspace{-4mm} &
      \includegraphics[width=0.12\textwidth]{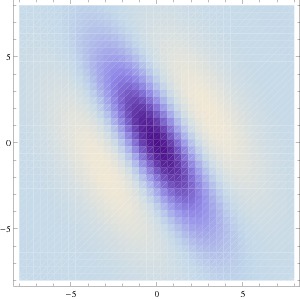} \hspace{-4mm} \\
    \end{tabular} 
  \end{center}
  \vspace{-3mm}
  \caption{Spatial receptive fields formed by affine Gaussian kernels and directional
    derivatives of these. The corresponding family of receptive fields
  is closed under general affine transformations of the spatial
  domain, including translations, rotations, scaling transformations and
  perspective foreshortening.}
  \label{fig-aff-Gaussian-ders}
\end{figure}

If we relax the requirement of rotational symmetry and relax it into the
requirement of mirror symmetry through the origin, then it follows
that the convolution kernels must instead be
{\em affine Gaussian kernels\/} \citep{Lin93-Dis}
\begin{equation}
  \label{eq-aff-gauss-kernel}
    T(x;\; s) = g(x;\; \Sigma)  = \frac{1}{2 \pi \sqrt{\det\Sigma}} e^{-x^T \Sigma^{-1} x/2}
\end{equation}
where $\Sigma$ denotes any symmetric positive semi-definite 
$2 \times 2$ matrix.
This affine scale-space concept is {\em closed\/} under affine
transformations,
meaning that if we for affine related images
\begin{equation}
  \label{eq-aff-trans-two-imgs-aff-scsp}
    f_L(\xi) = f_R(\eta) \quad \mbox{where} \quad \eta = A \, \xi + b.
\end{equation}
define corresponding scale-space representations according to
\begin{equation}
    L(\cdot;\; \Sigma_L) = g(\cdot;\; \Sigma_L) * f_L(\cdot), \quad
       R(\cdot;\; \Sigma_R) = g(\cdot;\; \Sigma_R) * f_R(\cdot)
\end{equation}
then these scale-space representations will be related according to \citep{Lin93-Dis,LG96-IVC}
\begin{equation}
  \label{eq-aff-transf-prop-aff-scsp}
    L(x;\; \Sigma_L) = R(y;\; \Sigma_R) \quad \mbox{where} \quad
     \Sigma_R = A \, \Sigma_L \, A^T
\quad \mbox{and} \quad y = A \, x + b.
\end{equation}
In other words, given that an image $f_L$ is affine transformed into
an image $f_R$ it will always be possible to find a transformation
between the scale parameters $s_L$ and $s_R$ in the two domains
that make it possible to match the corresponding derived 
 internal representations $L(\cdot;\; s_L)$ and $R(\cdot;\, s_R)$.
Figure~\ref{fig-aff-Gaussian-ders} shows a few examples of such
kernels in different directions with the covariance matrix
parameterized according to 
\begin{align}
  \label{eq-aff-cov-mat-2D}
 \begin{split}
  \Sigma & =
  \left(
    \begin{array}{ccc}
      \lambda_1 \cos^2 \theta + \lambda_2 \sin^2 \theta \quad
        &  (\lambda_1 - \lambda_2) \cos \theta \, \sin \theta 
        \\
      (\lambda_1 - \lambda_2) \cos \theta \, \sin \theta \quad
        & \lambda_1 \sin^2 \theta + \lambda_2 \cos^2 \theta
    \end{array}
  \right)
  \end{split}
\end{align}
with $\lambda_1$ and $\lambda_2$ denoting the eigenvalues and $\theta$
the orientation. Directional derivatives of these kernels can in turn
be obtained from linear combinations of partial derivative operators
according to
\begin{equation}
  \label{eq-dir-der-order-N}
  \partial_{\varphi^m} L 
  = (\cos \varphi \, \partial_{x_1} + \sin \varphi \, \partial_{x_2})^m L
  = \sum_{k=0}^m
      \left(
        \begin{array}{c}
          m \\
          k
        \end{array}
      \right)
      \cos^k \varphi \, \sin^{m-k} \varphi \, L_{x_1^k x_2^{m-k}}.
\end{equation}
With respect to biological vision, these kernels can be used for
modelling receptive fields that are oriented in the spatial domain,
as will be described in connection with
equation~(\ref{eq-orient-spat-rec-field-aff-gauss})
in section~\ref{sec-model-biol-rec-fields}.
For computer vision they can be used for computing 
{\em affine invariant image descriptors\/} for {\em e.g.\/}
cues to surface shape, image-based matching and recognition
\citep{Lin93-Dis,LG96-IVC,Bau00-CVPR,MikSch04-IJCV,TuyGoo04-IJCV,LazSchPon05-PAMI,RotLazSchPon06-IJCV}.

\paragraph{Note on receptive fields formed from derivatives of the
  convolution kernels.}

Due to the linearity of the differential equation
(\ref{eq-nec-diff-eq-spat-rec-fields}), which has been derived by necessity
from the structural requirements, it follows that also the result of applying a
linear operator ${\cal D}$ to the solution $L$ will also satisfy the
differential equation, however, with a different initial condition
\begin{equation}
   \label{eq-init-cond-scsp-der}
    \lim_{s \downarrow 0} ({\cal D} L)(\cdot;\; s) =  {\cal D} f.
\end{equation}
The result of applying a linear operator ${\cal D}$ to the scale-space
representation $L$ will therefore satisfy the above mentioned structural
requirements of linearity, shift invariance, the weaker form of
rotational invariance at the group level and non-enhancement of local extrema,
with the semi-group structure (\ref{eq-semi-group}) replaced by
the cascade property
\begin{equation}
  ({\cal D} L)(\cdot;\; s_2) 
  = T(\cdot;\; s_2-s_1) * ({\cal D} L)(\cdot;\; s_1).
\end{equation}
Then, one may ask if any linear operator ${\cal D}$ would be
reasonable?
From the requirement of scale invariance, however, if follows that that
the operator ${\cal D}$ must not be allowed to have non-infinitesimal
support, since a non-infinitesimal support $s_0 > 0$ would violate
the requirement of self-similarity over scale (\ref{eq-self-sim-req})
and it would not be possible to perform image measurements at a scale
level lower than $s_0$. 
Thus, any receptive field operator derived from the scale-space representation
in a manner compatible with the structural arguments must correspond
to local derivatives. In the illustrations above, partial derivatives and
directional derivatives up to order two have been shown.

For directional derivatives that have been derived from elongated kernels whose
underlying zero-order convolution kernels are not rotationally
symmetric, it should be noted that we have aligned the directions of the directional
derivative operators to the orientations of the underlying kernels.
A structural motivation for making such an alignment can be obtained
from a requirement of a weaker form of rotational symmetry at the
group level.
If we would like the family of receptive fields to be rotationally
symmetric as a group, then it is natural to require the
directional derivative operators to be transformed in a similar way as
the underlying kernels.

Receptive fields in terms of derivatives of the convolution kernels
derived by necessity do also have additional advantages if one adds a
further structural requirement of invariance under additive
intensity transformations $f(x) \mapsto f(x) + C$.
A zero-order receptive field will be affected by such an intensity
transformation, whereas higher order derivatives are invariant under
additive intensity transformations.
As will be described in section~\ref{sec-inv-illum-var}, this form of invariance has a
particularly interesting physical interpretation with regard to a logarithmic
intensity scale.

\subsection{Spatio-temporal image data}

For spatio-temporal image data $f(x, t)$ defined on a 2+1-D
spatio-temporal domain with $(x, t) = (x_1, x_2, t)$ it is natural to
inherit the symmetry requirements over the spatial domain.
In addition, the following structural requirements can be imposed
motivated by the special nature of time and space-time.

\paragraph{Galilean covariance.}

For time-dependent spatio-temporal image data, we may have 
{\em relative motions\/} between objects in the world and the
observer, where a constant velocity translational motion can be 
modelled by a  {\em Galilean transformation\/}
\begin{equation}
  \label{eq-def-gal-transf}
    f' = {\cal G}_v \, f  \quad \mbox{corresponding to} \quad
     f'(x', t') = f(x, t) \quad \mbox{with} \quad x' = x + v \, t.
\end{equation}
To enable a consistent visual interpretation under
different relative motions, it is natural to require that it should be
possible to transform internal representations 
$L(\cdot, \cdot;\; s)$ that are computed from spatio-temporal image
data under different relative motions
\begin{equation}
   L'(x', t';\; s') = L(x, t;\; s) \quad \mbox{corresponding to} \quad
         {\cal T}_{G_v(s)} \, {\cal G}_v \, f = {\cal G}_v \, {\cal T}_s \, f.
\end{equation}
Such a property is referred to as {\em Galilean covariance\/}.

\paragraph{Temporal causality.}

For a vision system that interacts in with the environment in a
real-time setting, a fundamental constraint on the convolution kernels
(the spatio-temporal receptive fields) is that they cannot access data
from the future.
Hence, they must be {\em time-causal\/} in the sense that convolution
kernel must be zero for any relative time moment that would imply
access to the future:
\begin{equation}
  \label{eq-def-time-caus}
    T(x, t;\; s) = 0 \quad \mbox{if} \quad t < 0.
\end{equation}

\paragraph{Time-recursivity.}

Another fundamental constraint on a real-time system is that it cannot
keep a record of everything that has happened in the past.
Hence, the computations must be based on a limited internal 
{\em temporal buffer\/} $M(x, t)$, which should provide:
\begin{itemize}
\item
    a sufficient record of past information and
\item
   sufficient information to update its internal state when new
   information arrives.
\end{itemize}
A particularly useful solution is to use the  internal 
representations $L$ at different temporal scales
also used as the memory buffer of the past.
In \citep[section~5.1.3, page 57]{Lin10-JMIV} it is shown that such a requirement can be
formalized by a time-recursive updating rule of the form
\begin{align}
    \begin{split}
      L(x, t_2; s_2, \tau) 
      & = 
        \int_{\xi \in \bbbr^N} 
        \int_{\zeta \geq 0} 
          U(x-\xi, t_2-t_1;\; s_2 - s_1, \tau, \zeta) \, L(\xi,t_1;\; s_1, \zeta) 
          \, d\zeta \, d\xi
   \end{split}\nonumber\\
    \begin{split}
      & \phantom{=}
      + \int_{\xi \in \bbbr^N} 
        \int_{u = t_1}^{t_2}
          B(x-\xi, t_2-u;\; s_2, \tau) \, f(\xi, u) \, d\xi \, du
    \end{split}\nonumber
\end{align}
which is required to hold for any pair of scale levels $s_2 \geq s_1$ and 
any two time moments $t_2 \geq t_1$, where
\begin{itemize}
\item
   the kernel $U$ updates the internal state,
\item
  the kernel $B$ incorporates new image data into the representation,
\item
  $\tau$ is the temporal scale and $\zeta$ an integration variable
  referring to internal temporal buffers at different temporal scales.
\end{itemize}

\paragraph{Non-enhancement of local extrema in a time-recursive
  setting.}

For a time-recursive spatio-temporal visual front-end it is also
natural to generalize the notion of non-enhancement of local extrema,
such that it is required to hold both with respect to increasing
spatial scales $s$ and evolution over time $t$.
Thus, if at some spatial scale $s_0$ and time moment $t_0$
a point  $(x_0, \tau_0)$ is a local maximum (minimum) for the mapping
\begin{equation}
    (x, \tau) \rightarrow L(x, t_0;\; s_0, \tau)
\end{equation}
  then for {\em every positive direction\/} $u = (u_1, \dots, u_N, u_{N+1})$ 
  in the $N+1$-dimensional 
  space spanned by $(s, t)$, the directional derivative 
  $({\cal D}_u L)(x, t;\; s, \tau)$ must satisfy 
\begin{align}
  \begin{split}
     ({\cal D}_u L)(x_0, t_0;\; s_0, \tau_0) \leq 0 \quad\quad \mbox{at any local maximum},
  \end{split}\\
  \begin{split}
     ({\cal D}_u L)(x_0, t_0;\; s_0, \tau_0) \geq 0 \quad\quad \mbox{at any local minimum}.
  \end{split}
\end{align}

\section{Necessity results concerning spatio-temporal receptive fields}

We shall now describe how these structural requirements restrict the
class of possible spatio-temporal receptive fields.

\subsection{Non-causal spatio-temporal receptive fields}
\label{sec-non-caus-spat-temp-scsp}

If one disregards the requirements of temporal causality and time
recursivity and instead requires
(i)~linearity, 
(ii)~shift invariance over space and time,
(iii)~semi-group property over spatial and temporal scales,
(iv)~sufficient regularity properties over space, time and
spatio-temporal scales and
(v)~non-enhancement of local extrema for a multi-parameter scale-space,
then it follows from \citep[theorem~5, page~42]{Lin10-JMIV}
that the scale-space representation over a 2+1-D spatio-temporal domain
must satisfy
\begin{equation}
    \partial_s L 
    = \frac{1}{2} \, \nabla_{(x,t)}^T \left( \Sigma_0 \nabla_{(x,t)} L \right) - \delta_0^T \, \nabla_{(x,t)} L
\end{equation}
for some $3 \times 3$ covariance matrix $\Sigma_0$ and 
some 3-D vector $\delta_0$
with $\nabla_{(x, t)} = (\partial_{x_1}, \partial_{x_2}, \partial_t)^T$.

\begin{figure}[hbtt]
   \begin{center}
   \begin{tabular}{c}
      \includegraphics[width=0.12\textwidth]{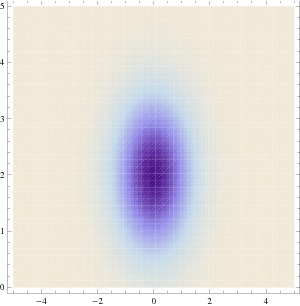} \hspace{-4mm} \\
    \end{tabular} 
  \end{center}
  \vspace{-10mm}
  \begin{center}
    \begin{tabular}{cc}
      \includegraphics[width=0.12\textwidth]{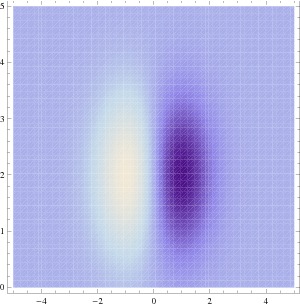} \hspace{-4mm} &
      \includegraphics[width=0.12\textwidth]{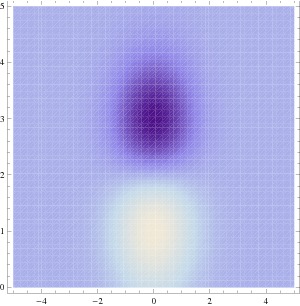} \hspace{-4mm} \\
    \end{tabular} 
  \end{center}
  \vspace{-10mm}
  \begin{center}
    \begin{tabular}{ccc}
      \includegraphics[width=0.12\textwidth]{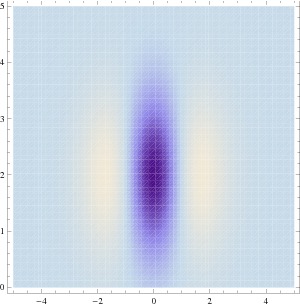} \hspace{-4mm} &
      \includegraphics[width=0.12\textwidth]{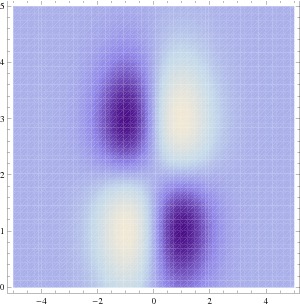} \hspace{-4mm} &
      \includegraphics[width=0.12\textwidth]{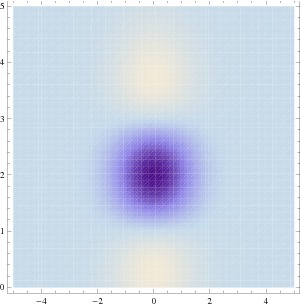} \hspace{-4mm} \\
    \end{tabular} 
  \end{center}
  \caption{Non-causal and space-time separable spatio-temporal
    receptive fields over 1+1-D space-time as generated by the
    Gaussian spatio-temporal scale-space model with $v = 0$.
    This family of receptive fields is closed under
    rescalings of the spatial and temporal dimensions. 
    (Horizontal axis: space $x$. Vertical axis: time~$t$.)}
  \label{fig-non-caus-sep-spat-temp-rec-fields}

  \bigskip

 \begin{center}
    \begin{tabular}{c}
      \includegraphics[width=0.12\textwidth]{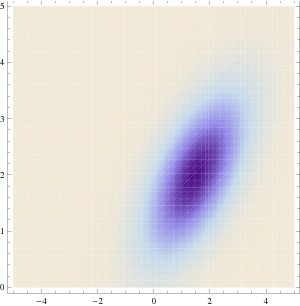} \hspace{-4mm} \\
    \end{tabular} 
  \end{center}
  \vspace{-10mm}
  \begin{center}
    \begin{tabular}{cc}
      \includegraphics[width=0.12\textwidth]{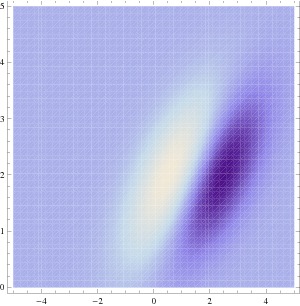} \hspace{-4mm} &
      \includegraphics[width=0.12\textwidth]{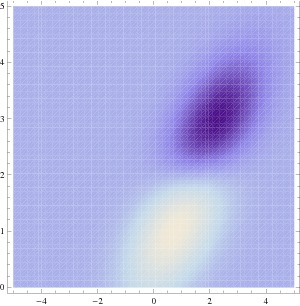} \hspace{-4mm} \\
    \end{tabular} 
  \end{center}
  \vspace{-10mm}
  \begin{center}
    \begin{tabular}{ccc}
      \includegraphics[width=0.12\textwidth]{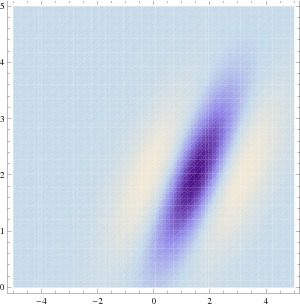} \hspace{-4mm} &
      \includegraphics[width=0.12\textwidth]{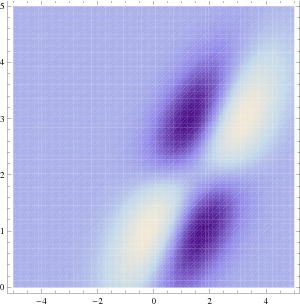} \hspace{-4mm} &
      \includegraphics[width=0.12\textwidth]{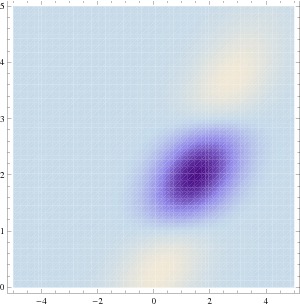} \hspace{-4mm} \\
    \end{tabular} 
  \end{center}
\caption{Non-causal and velocity-adapted spatio-temporal
    receptive fields over 1+1-D space-time as generated by the
    Gaussian spatio-temporal scale-space model for a non-zero image
    velocity $v$. This family of receptive fields is closed under
    rescalings of the spatial and temporal dimensions as well as
    Galilean transformations.
    (Horizontal axis: space $x$. Vertical axis: time $t$.)}
  \label{fig-non-caus-vel-adapt-spat-temp-rec-fields}
\end{figure}

In terms of convolution kernels, the zero-order receptive fields will
then be {\em spatio-temporal Gaussian kernels\/}
\begin{equation}
  \label{eq-gauss-spat-temp-kernel}
   g(p;\; \Sigma_s, \delta_s) = \frac{1}{(2 \pi)^{3/2} \sqrt{\det \Sigma_s}} \, e^{- {(p - \delta_s)^T \Sigma_s^{-1} (p - \delta_s)}/{2s}}
\end{equation}
with $p = (x, t)^T =  (x_1, x_2, t)^T$,
      \begin{align}
         \begin{split}
           \label{eq-spat-temp-cov-matrix-2+1-D-space-time}
           \Sigma_s & = 
           \left(
             \begin{array}{ccc}
               \lambda_1 \cos^2 \theta + \lambda_2 \sin^2 \theta + v_1^2 \lambda_t
                 &  (\lambda_2 - \lambda_1) \cos \theta \, \sin \theta + v_1 v_2 \lambda_t
                 & v_1 \lambda_t \\
               (\lambda_2 - \lambda_1) \cos \theta \, \sin \theta + v_1 v_2 \lambda_t
                 & \lambda_1 \sin^2 \theta + \lambda_2 \cos^2 \theta + v_2^2 \lambda_t
                 & v_2 \lambda_t \\
               v_1 \lambda_t & v_2 \lambda_t & \lambda_t
             \end{array}
           \right)
        \end{split}\\
        \begin{split}
          \delta_s & = 
          \left(
            \begin{array}{c}
               v_1 t \\
               v_2 t \\
               \delta
            \end{array}
          \right) 
         \end{split}
       \end{align}
where
(i)~$\lambda_1$, $\lambda_2$ and $\theta$ determine the 
       {\em spatial extent\/},
(ii)~$\lambda_t$ determines the {\em temporal extent\/},
(iii)~$v = (v_1, v_2)$ denotes the {\em image velocity\/}  and
(iv)~$\delta$ represents a {\em temporal delay\/}.
From the corresponding {\em Gaussian spatio-temporal scale-space\/}
\begin{equation}
   L(x, t;\; \Sigma_{space}, v, \tau)  = (g(\cdot, \cdot;\; \Sigma_{space}, v, \tau) * f(\cdot, \cdot))(x, t)
\end{equation}
spatio-temporal derivatives can then be defined according to
\begin{equation}
  \label{eq-spat-temp-ders}
    L_{x^{\alpha} t^{\beta}}(x, t;\; \Sigma_{space}, v, \tau)  =
    (\partial_{x^{\alpha} t^{\beta}} L)(x, t;\; \Sigma_{space}, v, \tau)
\end{equation}
with corresponding {\em velocity-adapted temporal derivatives\/}
\begin{equation}
  \label{eq-vel-adapt-spat-temp-ders}
    \partial_{\bar{t}} = v^T \nabla_x + \partial_t = v_1 \, \partial_{x_1} + v_2 \, \partial_{x_2} + \partial_t
\end{equation}
as illustrated in
figure~\ref{fig-non-caus-sep-spat-temp-rec-fields} and 
figure~\ref{fig-non-caus-vel-adapt-spat-temp-rec-fields} for the case
of a 1+1-D space-time.

Motivated by the requirement of Galilean covariance, it is natural to
align the directions $v$ in space-time for which these velocity-adapted
spatio-temporal derivatives are computed to the
velocity values used in the underlying zero-order spatio-temporal
kernels, since the resulting velocity-adapted spatio-temporal
derivatives will then be Galilean covariant.
Such receptive fields can be used for modelling spatio-temporal
receptive fields in biological vision \citep{Lin97-ICSSTCV,YouLesMey01-SV,YouLes01-SV,Lin10-JMIV}
and for computing spatio-temporal image
features and Galilean invariant image descriptors for 
spatio-temporal recognition in computer vision 
\citep{LapLin03-ICCV,LapLin03-IVC,LapLin04-ECCVWS,LapCapSchLin07-CVIU,WilTuyGoo08-ECCV}.

\paragraph{Transformation property under Galilean transformations.}

Under a Galilean transformation of space-time
(\ref{eq-def-gal-transf}), in matrix form written
\begin{equation}
  p' = G_v \, p 
  \quad\quad \mbox{corresponding to} \quad\quad
 \left(
    \begin{array}{c}
       x_1' \\
       x_2' \\
       t'
    \end{array}
  \right)
  =
  \left(
    \begin{array}{ccc}
       1 & 0 & v_1 \\
       0 & 1 & v_2 \\
       0 & 0 & 1 \\
     \end{array}
  \right)
 \left(
    \begin{array}{c}
       x_1 \\
       x_2 \\
       t
    \end{array}
  \right),
\end{equation}
the corresponding Gaussian spatio-temporal representations are related
in an algebraically similar way 
(\ref{eq-aff-trans-two-imgs-aff-scsp}--\ref{eq-aff-transf-prop-aff-scsp})
as the affine Gaussian scale-space with the affine transformation
matrix $A$ replaced by a Galilean transformation matrix $G_v$.
In other words, if two spatio-temporal image patterns $f_L$ and $f_R$
are related by a Galilean transformation encompassing a 
translation $\Delta p = (\Delta x_1, \Delta x_2, \Delta t)$ in space-time
\begin{equation}
   f_L(\xi) = f_R(\eta) \quad \mbox{where} \quad \eta = G_v \, \xi +
   \Delta p
\end{equation}
and if corresponding spatio-temporal scale-space representations are defined according to
\begin{equation}
    L(\cdot;\; \Sigma_L) = g(\cdot;\; \Sigma_L) * f_L(\cdot), \quad
       R(\cdot;\; \Sigma_R) = g(\cdot;\; \Sigma_R) * f_R(\cdot)
\end{equation}
for general spatio-temporal covariance matrices $\Sigma_L$ and
$\Sigma_R$ of the form
(\ref{eq-spat-temp-cov-matrix-2+1-D-space-time}),
then these spatio-temporal scale-space representations will be related according to 
\begin{equation}
   L(x;\; \Sigma_L) = R(y;\; \Sigma_R) 
   \quad \mbox{where} \quad \Sigma_R = G_v \, \Sigma_L \, G_v^T
   \quad \mbox{and} \quad y = G_v \, x + \Delta p.
\end{equation}
Given two spatio-temporal image patterns that are related by a
Galilean transformation, such as arising when an object is observed
with different relative motion between the object and the viewing
direction of the observer, it will therefore be possible to perfectly
match the spatio-temporal receptive field responses computed from the
different spatio-temporal image patterns.
Such a perfect matching would, however, not be possible without
velocity adaptation, {\em i.e.\/}, if the spatio-temporal receptive
fields would be computed using space-time separable receptive fields only.

\begin{figure}[!b]
  \begin{center}
    \begin{tabular}{c}
      \includegraphics[width=0.12\textwidth]{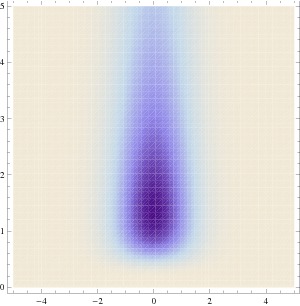} \hspace{-4mm} \\
    \end{tabular} 
  \end{center}
  \vspace{-8mm}
  \begin{center}
    \begin{tabular}{cc}
      \includegraphics[width=0.12\textwidth]{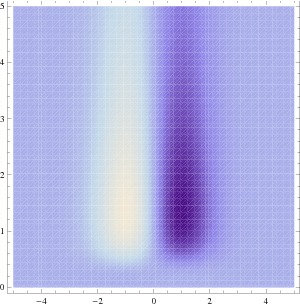} \hspace{-4mm} &
      \includegraphics[width=0.12\textwidth]{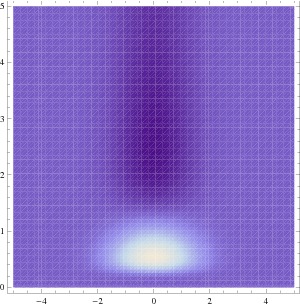} \hspace{-4mm} \\
    \end{tabular} 
  \end{center}
  \vspace{-8mm}
  \begin{center}
    \begin{tabular}{ccc}
      \includegraphics[width=0.12\textwidth]{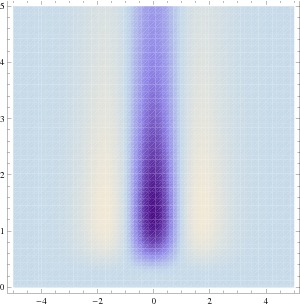} \hspace{-4mm} &
      \includegraphics[width=0.12\textwidth]{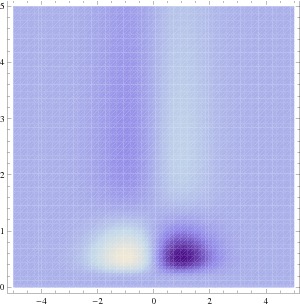} \hspace{-4mm} &
      \includegraphics[width=0.12\textwidth]{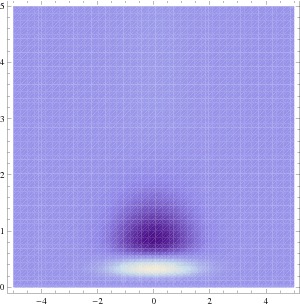} \hspace{-4mm} \\
    \end{tabular} 
  \end{center}
  \caption{Time-causal and space-time separable spatio-temporal
    receptive fields over a 1+1-D space-time as generated by the
    time-causal spatio-temporal scale-space model with $v = 0$.
    This family of receptive fields is closed under
    rescalings of the spatial and temporal dimensions.
    (Horizontal axis: space $x$. Vertical axis: time $t$.)}
  \label{fig-time-caus-sep-spat-temp-rec-fields-1+1-D}
  \begin{center}
    \begin{tabular}{c}
      \includegraphics[width=0.12\textwidth]{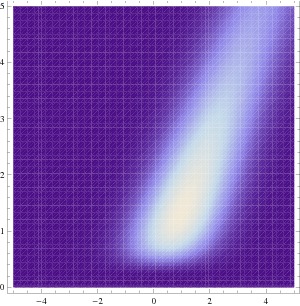} \hspace{-4mm} \\
    \end{tabular} 
  \end{center}
  \vspace{-8mm}
  \begin{center}
    \begin{tabular}{cc}
      \includegraphics[width=0.12\textwidth]{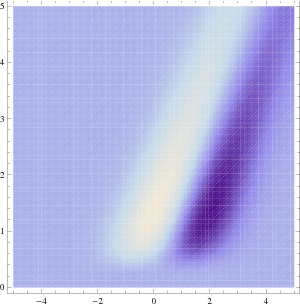} \hspace{-4mm} &
      \includegraphics[width=0.12\textwidth]{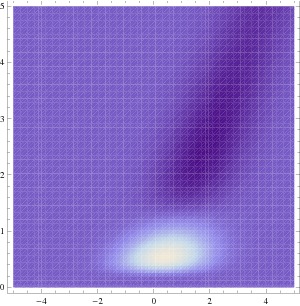} \hspace{-4mm} \\
    \end{tabular} 
  \end{center}
  \vspace{-8mm}
  \begin{center}
    \begin{tabular}{ccc}
      \includegraphics[width=0.12\textwidth]{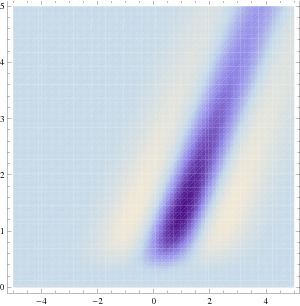} \hspace{-4mm} &
      \includegraphics[width=0.12\textwidth]{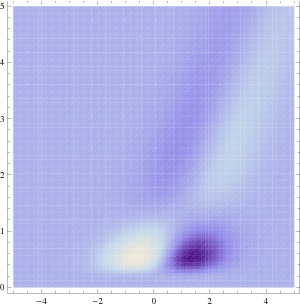} \hspace{-4mm} &
      \includegraphics[width=0.12\textwidth]{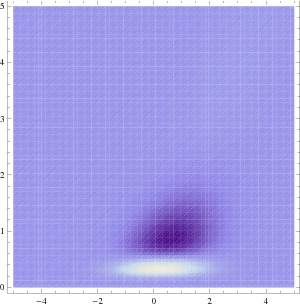} \hspace{-4mm} \\
    \end{tabular} 
  \end{center}
  \caption{Time-causal and velocity-adapted spatio-temporal
    receptive fields over a 1+1-D space-time as generated by the
    time-causal spatio-temporal scale-space model with $v = 0$.
    This family of receptive fields is closed under
    rescalings of the spatial and temporal dimensions as well as 
    Galilean transformations.
    (Horizontal axis: space $x$. Vertical axis: time $t$.)}
  \label{fig-time-caus-vel-adapt-spat-temp-rec-fields-1+1-D}
\end{figure}

\subsection{Time-causal spatio-temporal receptive fields}
\label{sec-time-caus-spat-temp-scsp}

If we on the other hand with regard to real-time biological vision
want to respect both temporal causality and temporal
recursivity, we obtain a different family of {\em time-causal\/} spatio-temporal receptive fields.
Given the requirements of
(i)~linearity,
(ii)~shift invariance over space and time,
(iii)~temporal causality,
(iv)~time-recursivity,
(v)~semi-group property over spatial scales $s$ and time $t$ 
      ${\cal T}_{s_1, t_1} \, {\cal T}_{s_2, t_2} = {\cal T}_{s_1+s_2, t_1+t_2}$,
(vi)~sufficient regularity properties over space, time and
spatio-temporal scales and
(vii)~non-enhancement of local extrema in a time-recursive setting%
\footnote{Concerning the relations between the non-causal
                spatio-temporal model in
                section~\ref{sec-non-caus-spat-temp-scsp}
                and the time-causal model in
                section~\ref{sec-time-caus-spat-temp-scsp}, 
                please note that requirement of non-enhancement of
                local extrema is formulated in different ways in the two cases:
(i)
                For the non-causal scale-space model, the condition about 
                non-enhancement condition is based on points that are
                local extrema with respect to both space $x$ and time
                $t$. At such points, a sign condition is imposed on
                the derivative in any positive direction over spatial
                scales $s$ and temporal scale $\tau$.
(ii)
                For the time-causal scale-space model, the notion of 
                local extrema is based on points that are local
                extrema with respect to space $x$ and the internal 
                temporal buffers at different temporal scales $\tau$. 
                At such points, a sign
                condition is imposed on the derivatives in the
                parameter space defined by the spatial scale parameters
                $s$ and time $t$.
                Thus, in addition to the restriction to time-causal
                convolution kernels (\protect\ref{eq-def-time-caus}) 
                the derivation of the time-causal scale-space model
                is also based on different structural requirements.},
then it
follows that the time-causal spatio-temporal scale-space must satisfy
the {\em system\/} of diffusion equations
\citep[equations~(88-89), page 52, theorem~17, page~78]{Lin10-JMIV}
\begin{align}
    \begin{split}
     \partial_s L 
        & = \frac{1}{2} \nabla_x^T (\Sigma \nabla_x L)
    \end{split}\\
    \begin{split}
     \partial_t L 
        & = - v^T \nabla_x L + \frac{1}{2} \partial_{\tau\tau} L
    \end{split}
\end{align}
for some $2 \times 2$ spatial covariance matrix $\Sigma$ and
some image velocity $v$
with $s$ denoting the {\em spatial scale\/}
and $\tau$ the {\em temporal scale\/}.
In terms of receptive fields, this spatio-temporal scale-space can
be computed by convolution kernels of the form
\begin{align}
   \begin{split}
       h(x, t;\; s, \tau;\; \Sigma, v)
         & = g(x - v t;\; s;\; \Sigma) \, \phi(t;\; \tau)
    \end{split}\nonumber\\
    \begin{split}
      \label{eq-time-caus-spat-temp-kernel}
      & = \frac{1}{2 \pi s \sqrt{\det \Sigma}} \,
           e^{-(x - v t)^T \Sigma^{-1} (x - v t)/2s} \,
           \frac{1}{\sqrt{2 \pi} \, t^{3/2}} \,
           \tau \, e^{-\tau^2/2t}
    \end{split}
\end{align}
where
\begin{itemize}
\item
    $g(x - v t;\; s;\; \Sigma)$ is a {\em velocity-adapted 2-D affine
    Gaussian kernel\/} with covariance matrix $\Sigma$ and
\item
  $\phi(t;\; \tau)$ is a {\em time-causal smoothing kernel over time\/} 
  with temporal scale parameter $\tau$.
\end{itemize}
From these kernels, spatio-temporal partial derivatives and
velocity-adapted derivatives can be computed in a corresponding manner
(\ref{eq-spat-temp-ders}) and (\ref{eq-vel-adapt-spat-temp-ders})
as for the Gaussian spatio-temporal scale-space concept;
see figure~\ref{fig-time-caus-sep-spat-temp-rec-fields-1+1-D} and
figure~\ref{fig-time-caus-vel-adapt-spat-temp-rec-fields-1+1-D} for
illustrations in the case of a 1+1-D space-time.


\section{Computational modelling of biological receptive fields}
\label{sec-model-biol-rec-fields}

An attractive property of the presented framework for early receptive
fields is that it generates receptive field profiles in good agreement
with receptive field profiles found by cell recordings in the retina,
LGN and V1 of higher mammals.
\citet{DeAngOhzFre95-TINS} and \citet{deAngAnz04-VisNeuroSci} present 
overviews of receptive fields in the {\em joint\/} space-time domain.
As outlined in \citep[section~6]{Lin10-JMIV},
the Gaussian and time-causal scale-space concepts presented here can be used for 
generating predictions of receptive field profiles that are 
qualitatively very similar to {\em all\/} the spatial and
spatio-temporal receptive fields presented in these surveys.

\subsection{LGN neurons}

\begin{figure}[!h]
   \begin{center}
     \begin{tabular}{cc}
        & {\small $\nabla^2 g(x, y;\; s)$} \\
       \includegraphics[width=0.61\textwidth]{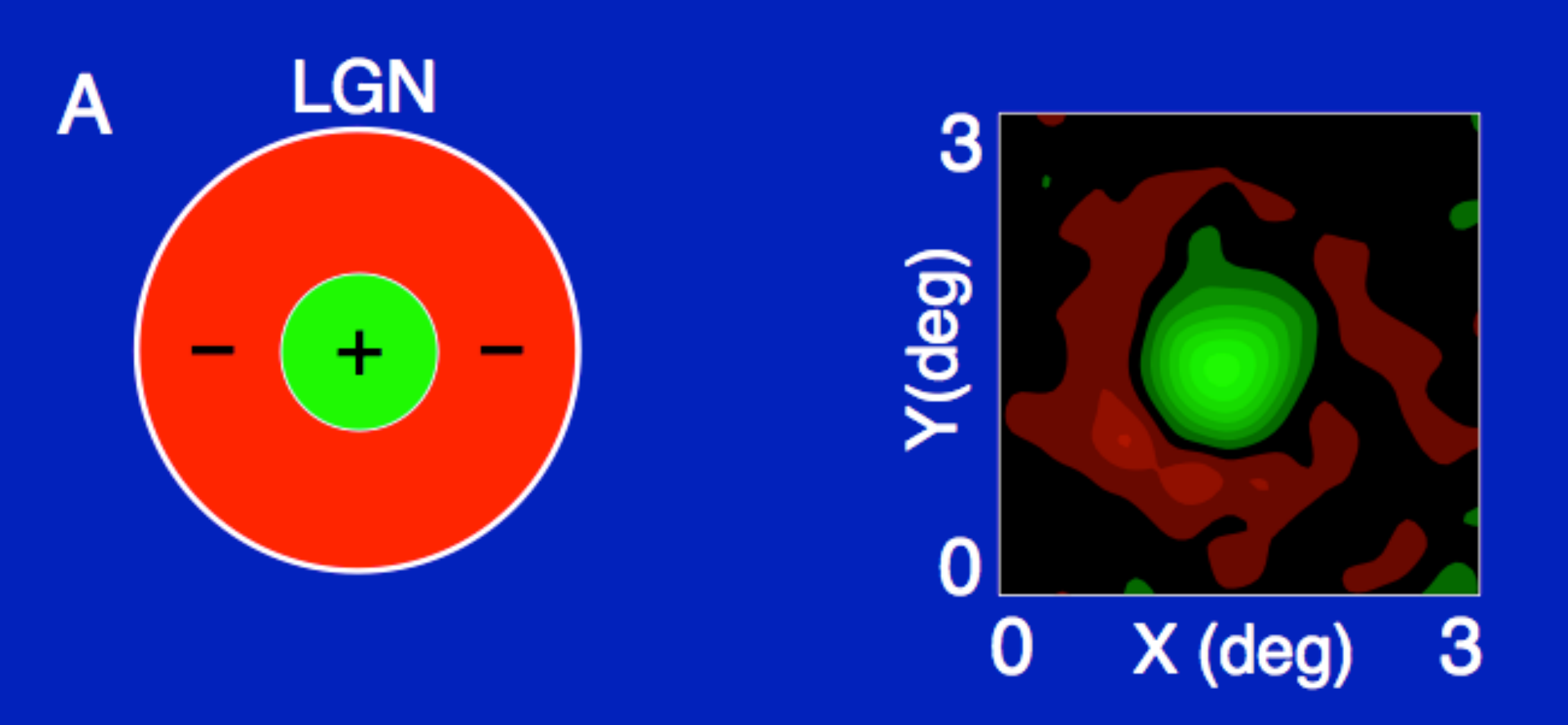}
       & \includegraphics[width=0.27\textwidth]{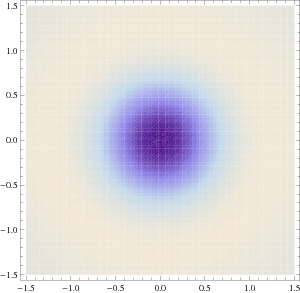}
     \end{tabular}
   \end{center}

  \caption{(left) Receptive fields in the LGN have approximately
               circular center-surround responses in the spatial
               domain, as reported by \protect\cite{DeAngOhzFre95-TINS}.
               (right)
               In terms of Gaussian derivatives, this spatial response 
               profile can be modelled by the Laplacian of the
               Gaussian $\nabla^2 g(x, y;\; s) 
            = (x^2 + y^2 - 2s)/(2 \pi s^3) \exp(-(x^2+y^2)/2s)$,
               here with $s = 0.4~\mbox{deg}^2$.}
  \label{fig-deang-tins-LGN-spat}
\end{figure}

In the LGN, most cells (i)~have approximately {\em circular-center surround\/} 
and most receptive fields are (ii)~{\em space-time separable\/} \citep{DeAngOhzFre95-TINS,deAngAnz04-VisNeuroSci}.
A corresponding idealized scale-space model for such receptive fields
can be expressed as
\begin{equation}
    h_{LGN}(x_1, x_2, t;\; s, \tau) = \pm (\partial_{x_1 x_1} + \partial_{x_2 x_2 }) \, g(x_1, x_2;\; s) \, \partial_{{t'}^n} \, h(t;\; \tau)
\end{equation}
where
\begin{itemize}
\item
    $\pm$ determines the  polarity (on-center/off-surround
     {\em vs:\/} off-center/on-surround),
\item
    $\partial_{x_1 x_1} + \partial_{x_2 x_2}$ denotes the spatial
    Laplacian operator,
\item
   $g(x_1, x_2;\; s)$ denotes a rotationally symmetric spatial
   Gaussian,
\item
  $\partial_{{t'}}$ denotes a temporal derivative operator with
  respect to a possibly self-similar transformation of time
  $t' = t^{\alpha}$ or $t' = \log t$ such that 
  $\partial_{{t'}} = t^{\kappa} \, \partial_t$ for some constant
  $\kappa$
  \cite[section~5.1, pages~59--61]{Lin10-JMIV},
\item
   $h(t;\; \tau)$ is a temporal smoothing kernel over time
  corresponding to the time-causal smoothing kernel 
  $\phi(t;\; \tau) = \tfrac{1}{\sqrt{2 \pi} \, t^{3/2}} \, \tau \, e^{-\tau^2/2t}$ 
  in (\ref{eq-time-caus-spat-temp-kernel}) or a
  non-causal time-shifted Gaussian kernel 
  $g(t;\; \tau, \delta) = \tfrac{1}{\sqrt{2 \pi \tau}} e^{-(t - \delta)^2/2 \tau}$ 
  according to (\ref{eq-gauss-spat-temp-kernel}),
\item
  $n$ is the order of temporal differentiation,
\item
   $s$ is the spatial scale parameter and
\item
   $\tau$ is the temporal scale parameter.
\end{itemize}
Figure~\ref{fig-deang-tins-LGN-spat} shows a comparison between 
the spatial component of a receptive field in the LGN with a Laplacian
of the Gaussian.
This model can also be used for modelling on-center/off-surround and
off-center/on-surround receptive fields in the retina.

Regarding the spatial domain, the model in terms of spatial
Laplacians of Gaussians
$(\partial_{x_1 x_1} + \partial_{x_2 x_2 }) \, g(x_1, x_2;\; s)$ 
is closely related to
differences of Gaussians, which have previously been shown to
be good approximation of the spatial variation of receptive fields in
the retina and the LGN \citep{Rod65-VisRes}.
This property follows from the fact that the
rotationally symmetric Gaussian satisfies the isotropic diffusion
equation
\begin{equation}
  \label{eq-expl-DoG-Laplace}
  \frac{1}{2} \nabla^2 L(x;\; t) 
  = \partial_t L(x;\; t) 
  \approx 
  \frac{L(x;\; t + \Delta t) - L(x;\; t)}{\Delta t}
  =  \frac{DOG(x;\; t, \Delta t)}{\Delta t}
\end{equation}
which implies that differences of Gaussians can be interpreted
as approximations of derivatives over scale and hence to Laplacian
responses.

\subsection{Simple cells in V1}

In V1 the receptive fields are generally different from the receptive
fields in the LGN in the sense that they are (i)~{\em oriented in the spatial
domain\/} and (ii)~{\em sensitive to specific stimulus velocities\/}
\citep{DeAngOhzFre95-TINS,deAngAnz04-VisNeuroSci}. 

\subsubsection{Spatial dependencies}

We can express a scale-space model for the {\em spatial component\/} of this
orientation dependency according to
\begin{equation}
  \label{eq-orient-spat-rec-field-aff-gauss}
    h_{space}(x_1, x_2;\; s) 
         = (\cos \varphi \, \partial_{x_1} + \sin \varphi \, \partial_{x_2})^m \, 
            g(x_1, x_2;\; \Sigma)
\end{equation}
where
\begin{itemize}
\item
   $\partial_{\varphi} = \cos \varphi \, \partial_{x_1} + \sin \varphi \, \partial_{x_2}$ 
   is a directional derivative operator,
\item
   $m$ is the order of spatial differentiation and
\item
   $g(x_1, x_2;\; \Sigma)$ is an affine Gaussian kernel with spatial
   covariance matrix $\Sigma$ as can be parameterized according to
   (\ref{eq-aff-cov-mat-2D})
\end{itemize}
where the direction $\varphi$ of the directional derivative operator
should preferably be aligned to the orientation $\theta$ of one of
the eigenvectors of $\Sigma$.

\begin{figure}[!h]
   \begin{center}
     \begin{tabular}{cc}
        & {\small $\partial_x g(x, y;\; \Sigma)$} \\
       \includegraphics[width=0.61\textwidth]{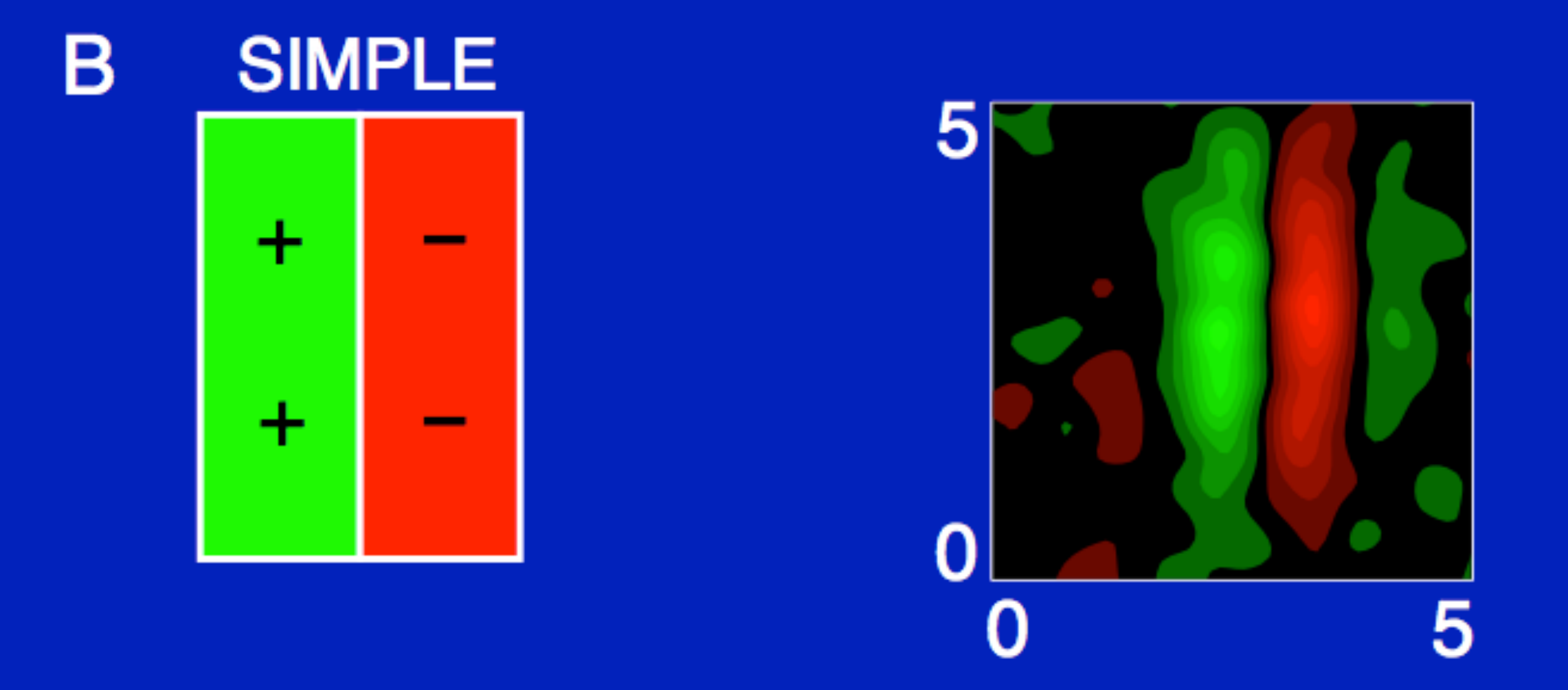}
       & \includegraphics[width=0.27\textwidth]{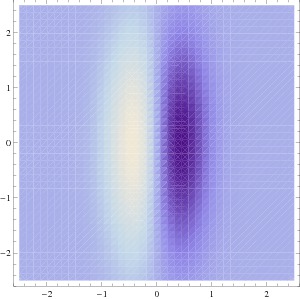}
     \end{tabular}
   \end{center}

  \caption{(left) Simple cells in the cerebral cortex do usually have
               strong directional preference in the spatial domain.
	       In terms of Gaussian derivatives, this spatial response
               can be modelled as a directional derivative 
               of an elongated affine Gaussian kernel, 
               as reported by \protect\cite{DeAngOhzFre95-TINS}.
               (right) First-order directional derivatives of anisotropic
               affine Gaussian kernels, here aligned to the
              coordinate directions  
             $\partial_x g(x, y;\; \Sigma) = 
              \partial_x g(x, y;\; \lambda_x, \lambda_y) 
              = - \frac{x}{\lambda_x} 1/(2 \pi \sqrt{\lambda_x \lambda_y}) 
                     \exp(-x^/2\lambda_x -y^2/2\lambda_y)$ 
             and here with $\lambda_x = 0.2~\mbox{deg}^2$ 
             and $\lambda_y = 2~\mbox{deg}^2$,
             can be used as a model for simple
             cells with a strong directional preference.}
  \label{fig-deang-simple-LGN-spat}
\end{figure}

In the specific case when the covariance matrix is proportional to a
unit matrix $\Sigma = s \, I$, with $s$ denoting the spatial scale
parameter, these directional derivatives correspond to regular
Gaussian derivatives as proposed as a model for spatial receptive
fields by \citet{KoeDoo87-BC} and \citet{KoeDoo92-PAMI}.
The use of non-isotropic covariance matrices do on the other hand
allow for a higher degree of orientation selectivity.
Moreover, by having a family of affine adapted kernels tuned to a
family of covariance matrices with different orientations and
different ratios between the scale parameters in the two directions, 
the family as a whole can represent affine covariance which makes
it possible to perfectly match corresponding receptive field responses
between different views obtained under variations of the viewing
direction in relation to the object.

\begin{figure}[!h]
   \begin{center}
     \begin{tabular}{cc}
       \includegraphics[width=0.42\textwidth]{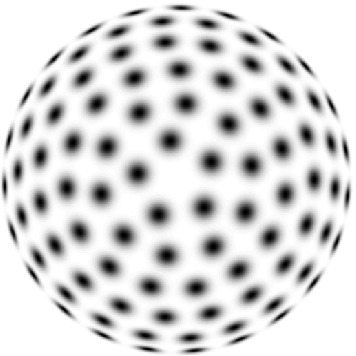} $\quad\quad$
       & \includegraphics[width=0.42\textwidth]{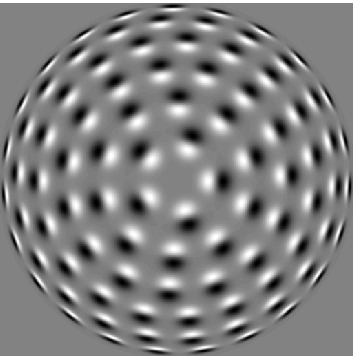}
     \end{tabular}
   \end{center}

  \caption{Affine Gaussian receptive fields generated for a set of
    covariance matrices $\Sigma$ that correspond to an approximately
    uniform distribution on a hemisphere in the 3-D environment, 
    which is then projected onto a 2-D image plane. 
    (left) Zero-order receptive fields. (right) First-order receptive fields.}
  \label{fig-aff-rec-fields-sphere}
\end{figure}

Figure~\ref{fig-aff-rec-fields-sphere} shows illustrations of
affine receptive fields of different orientations and degrees of
elongation as they arise if we assume that the set of all 
3-D objects in the world have an approximately uniform distribution 
of surface orientations in 3-D space and if we furthermore assume
that we observe these objects from a uniform distribution of viewing
directions that are not directly coupled to properties of the objects.

This idealized model of elongated receptive fields can also be extended to recurrent intracortical feedback
mechanisms as formulated by \citet{SomeNelSur95-JNeurSci} and
\citet{SomSha97-CurrOpNeurBiol}
by starting from the equivalent formulation in terms of the
non-isotropic diffusion equation  (\ref{eq-nec-diff-eq-spat-rec-fields})
\begin{equation}
    \partial_s L 
    = \frac{1}{2} \nabla_x^T \left( \Sigma_0 \nabla_x L \right) 
\end{equation} 
with the covariance matrix $\Sigma_0$ locally adapted%
\footnote{By the use of locally adapted feedback, the resulting
  evolution equation does not obey the original linearity and
  shift-invariance (homogeneity) requirements used for deriving the
  idealized affine Gaussian receptive field model, if the
  covariance matrices $\Sigma_0$ are determined from a
  properties of the image data that are determined in a non-linear
  way. For a fixed set of covariance matrices $\Sigma_0$ at any image
  point, the evolution equation will still be linear and will
  specifically obey non-enhancement of local extrema.
  In this respect, the resulting model could be regarded as a simplest
  form of non-linear extension of the idealized receptive field
  model.}
 to the statistics of image data in a neighbourhood of each image point;
see \citet{Wei98-book} and \citet{AL00-IP} for applications of this idea to the
enhancement of local directional image structures in computer vision.

\paragraph{Relations to modelling by Gabor functions:}

Gabor functions have been frequently used for modelling spatial
receptive fields \citep{Mar80-JOSA,JonPal87a,JonPal87b}, motivated by
their property of minimizing the uncertainty relation.
This motivation can, however, be questioned on both theoretical and
empirical grounds. 
\citet{StoWil90-JOSA} argue that (i)~only complex-valued Gabor
functions that cannot describe single receptive field minimize the
uncertainty relation, (ii)~the real functions that minimize this
relation are Gaussian derivatives rather than Gabor functions and
(iii)~comparisons among Gabor and alternative fits to both
psychophysical and physiological data have shown that in many cases
other functions (including Gaussian derivatives) provide better fits
than Gabor functions do.

Conceptually, the ripples of the Gabor functions, which are given by
complex sine waves, are related to the ripples of Gaussian
derivatives, which are given by Hermite functions.
A Gabor function, however, requires the specification of a scale parameter and a
frequency, whereas a Gaussian derivative requires a scale parameter
and the order of differentiation.
With the Gaussian derivative model, receptive fields of different
orders can be mutually related by derivative operations, 
and be computed from each other by nearest-neighbour operations. 
The zero-order receptive fields as well as the derivative based
receptive fields can be modelled by diffusion equations, 
and can therefore be implemented by computations between 
neighbouring computational units.

In relation to invariance properties, the family of affine Gaussian
kernels is closed under affine image deformations, whereas the family
of Gabor functions obtained by multiplying rotationally symmetric
Gaussians with sine and cosine waves is not closed under affine image
deformations. This means that it is not possibly to compute truly
affine invariant image representations from such Gabor functions.
Instead, given a pair of images that are related by a non-uniform
image deformation, the lack of affine covariance implies that there 
will be a systematic bias in image representations derived from such
Gabor functions, corresponding to the difference between the
backprojected Gabor functions in the two image domains.
If using receptive profiles defined from directional derivatives of
affine Gaussian kernels, it will on the other hand be possible to
compute affine invariant image representations.

In this respect, the Gaussian derivative model can be regarded as simpler, it can be
related to image measurements by differential geometry, be derived
axiomatically from symmetry principles, be computed from a minimal
set of connections and allows for provable invariance properties under
non-uniform (affine) image deformations.
\citet{You87-SV} has more generally shown how spatial receptive fields
in cats and monkeys can be well modelled by Gaussian derivatives up to order four. 

\begin{figure}[hbtp]
  \begin{center}
      \includegraphics[width=0.92\textwidth]{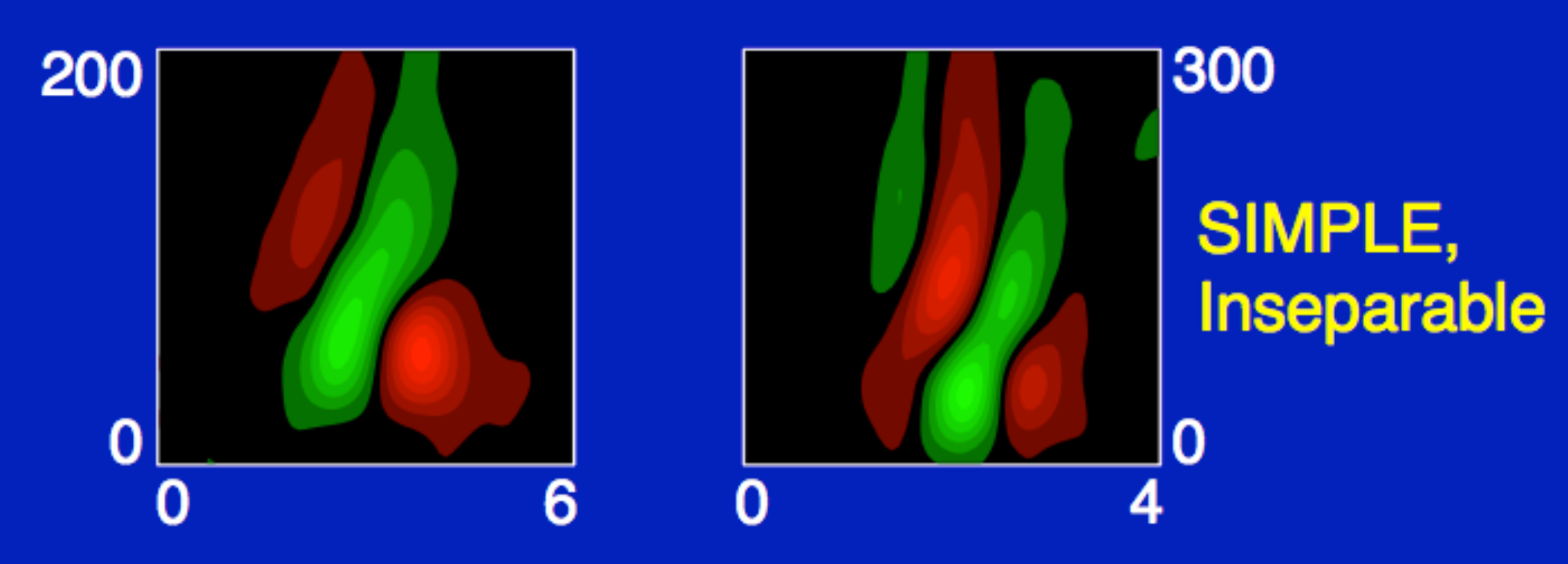} 
  \end{center}
\bigskip

  \begin{center}
    \begin{tabular}{cc}
      \hspace{-2.4cm} $g_{xx}(x, t;\; s, \tau, \delta)$
      & \hspace{0.5cm}  $-g_{xxx}(x, t;\; s, \tau, \delta)$ \\
      \hspace{-2.4cm} \includegraphics[width=0.27\textwidth]{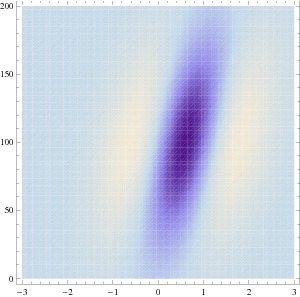}
      & \hspace{0.5cm} \includegraphics[width=0.27\textwidth]{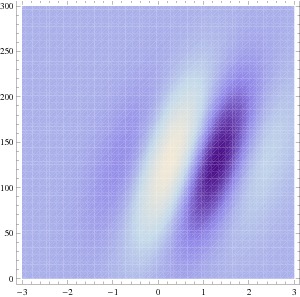}
    \end{tabular}

    \begin{tabular}{cc}
      \hspace{-2.4cm} $h_{xx}(x, t;\; s, \tau)$
      & \hspace{0.5cm} $-h_{xxx}(x, t;\; s, \tau)$ \\
      \hspace{-2.4cm} \includegraphics[width=0.27\textwidth]{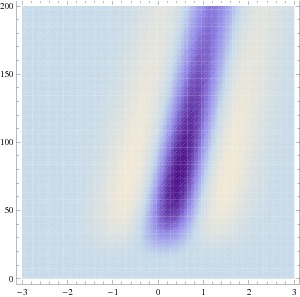}
      & \hspace{0.5cm} \includegraphics[width=0.27\textwidth]{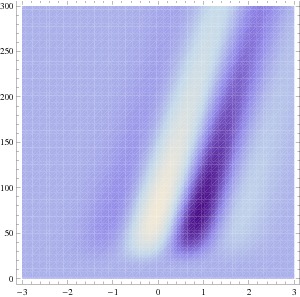}
    \end{tabular}
  \end{center}

\caption{(top row) Examples of non-separable spatio-temporal receptive field profiles in striate
           cortex as reported by \protect\cite{DeAngOhzFre95-TINS}:
           (top left) a receptive reminiscent of a second-order derivative
               in tilted space-time
               (compare with the left column in
                figure~\ref{fig-veladapt-space-time-kernel-1+1-D})
           (top right) a receptive reminiscent of a third-order derivative
               in tilted space-time
	       (compare with the right column in
                figure~\ref{fig-veladapt-space-time-kernel-1+1-D}).
            (middle and bottom rows) Non-separable spatio-temporal receptive fields obtained by
           applying velocity-adapted second- and third-order 
           derivative operations
           in space-time to spatio-temporal smoothing kernels
           generated by the spatio-temporal scale-space concept.
           (middle left) Gaussian spatio-temporal kernel
           $g_{xx}(x, t;\; s, \tau, v, \delta)$ with
           $s = 0.5~\mbox{deg}^2, \tau = 50^2~\mbox{ms}^2, 
            v = 0.006~\mbox{deg/ms}, \delta = 100~\mbox{ms}$.
            (middle right) Gaussian spatio-temporal kernel
           $g_{xxx}(x, t;\; s, \tau, v, \delta)$ with
           $s = 0.5~\mbox{deg}^2, \tau = 60^2~\mbox{ms}^2, 
             v = 0.006~\mbox{deg/ms}, \delta = 130~\mbox{ms}$.
           (lower left) Time-causal spatio-temporal kernel
           $h_{xx}(x, t;\; s, \tau, v)$ with
           $s = 0.4~\mbox{deg}^2, \tau = 15~\mbox{ms}^{1/2}, 
             v = 0.006~\mbox{deg/ms}$.
           (lower right) Time-causal spatio-temporal kernel
           $h_{xxx}(x, t;\; s, \tau, v)$ with
           $s = 0.4~\mbox{deg}^2, \tau = 15~\mbox{ms}^{1/2}, 
             v = 0.006~\mbox{deg/ms}$.
           (Horizontal dimension: space $x$. Vertical dimension: time~$t$.)}
  \label{fig-veladapt-space-time-kernel-1+1-D}
\end{figure}

\subsubsection{Spatio-temporal dependencies}

To model spatio-temporal receptive fields in the {\em joint space-time
domain\/}, we can then state scale-space models of simple cells in V1 
using either
\begin{itemize}
\item
   {\em non-causal Gaussian spatio-temporal derivative kernels\/}
   \begin{equation}
       h_{Gaussian}(x_1, x_2, t;\; s, \tau, v, \delta)
       = (\partial_{\varphi}^{\alpha_1} \, \partial_{\orth \varphi}^{\alpha_2}  \, \partial_{\bar{t}^n} g)(x_1, x_2, t;\; s, \tau, v, \delta)
   \end{equation}
\item
     {\em time-causal spatio-temporal derivative kernels\/}
     \begin{equation}
        h_{time-causal}(x_1, x_2, t;\; s, \tau, v) 
        = (\partial_{x_1^{\alpha_1} x_2^{\alpha_2}} \partial_{\bar{t}^n} h)(x_1, x_2, t;\; s, \tau, v)
     \end{equation}
\end{itemize}
with the non-causal Gaussian spatio-temporal kernels 
$g(x_1, x_2, t;\; s, \tau, v, \delta)$ according to
(\ref{eq-gauss-spat-temp-kernel}),
the time-causal spatio-temporal kernels 
$h(x_1, x_2, t;\; s, \tau, v)$ according to
(\ref{eq-time-caus-spat-temp-kernel})
and spatio-temporal derivatives 
$\partial_{x_1^{\alpha_1}  x_2^{\alpha_2}} \partial_{t^{\beta}}$ 
or velocity-adapted derivatives 
$\partial_{x_1^{\alpha_1}  x_2^{\alpha_2}} \partial_{\bar{t}^{\beta}}$  
of these according to
(\ref{eq-spat-temp-ders}) and (\ref{eq-vel-adapt-spat-temp-ders}).

For a general orientation of receptive fields with respect to the
spatial coordinate systems, the receptive fields in these scale-space models can be jointly
described in the form
\begin{align}
  \begin{split}
       h_{simple-cell}(x_1, x_2, t;\; s, \tau, v, \Sigma)  
       = & (\cos \varphi \, \partial_{x_1} + \sin \varphi \, \partial_{x_2})^{\alpha_1}
              (\sin \varphi \, \partial_{x_1} - \cos \varphi \, \partial_{x_2})^{\alpha_2}
  \end{split}\nonumber\\
  \begin{split}
         & (v_1 \, \partial_{x_1} + v_2 \, \partial_{x_2} + \partial_t)^n
    \end{split}\nonumber\\
  \begin{split}
        g(x_1 - v_1 t, x_2 - v_2 t;\; s \, \Sigma) \, h(t;\; \tau)
   \end{split}
\end{align}
  where
  \begin{itemize}
  \item
      $\partial_{\varphi} = \cos \varphi \, \partial_{x_1} + \sin \varphi \, \partial_{x_2}$ 
     and
     $\partial_{\orth \varphi} = \sin \varphi \, \partial_{x_1} - \cos \varphi \, \partial_{x_2}$
     denote spatial directional derivative operators according to
     (\ref{eq-dir-der-order-N}) in two orthogonal directions $\varphi$
     and $\orth \varphi$,
  \item
     $\alpha_1 \geq 0$ and $\alpha_2 \geq 0$ denote the orders of differentiation in the two
     orthogonal directions in the spatial domain with the overall
     spatial order of differentiation $m = \alpha_1 + \alpha_2$,
  \item 
    $v_1 \, \partial_{x_1} + v_2 \, \partial_{x_2} + \partial_t$ denotes a
    velocity-adapted temporal derivative operator,
  \item
   $v = (v_1, v_2)$ denotes the image velocity,
  \item
    $n$ denotes the order of temporal differentiation,
  \item
     $g(x_1 - v_1 t, x_2 - v_2 t;\; \Sigma)$ denotes a spatial affine Gaussian kernel
     according to (\ref{eq-aff-gauss-kernel}) that translates with image
     velocity $v = (v_1, v_2)$ in space-time,
  \item
    $\Sigma$ denotes a spatial covariance matrix that can be
    parameterized by two eigenvalues $\lambda_1$ and $\lambda_2$ as
    well as an orientation $\theta$ of the form
    (\ref{eq-aff-cov-mat-2D}),
\item
   $h(t;\; \tau)$ is a temporal smoothing kernel over time
  corresponding to the time-causal smoothing kernel 
  $\phi(t;\; \tau) = \tfrac{1}{\sqrt{2 \pi} \, t^{3/2}} \, \tau \, e^{-\tau^2/2t}$ 
  in (\ref{eq-time-caus-spat-temp-kernel}) or a
  non-causal time-shifted Gaussian kernel 
  $g(t;\; \tau, \delta) = \tfrac{1}{\sqrt{2 \pi \tau}} e^{-(t - \delta)^2/2 \tau}$ 
  according to (\ref{eq-gauss-spat-temp-kernel}),
  \item
    $s$ denotes the spatial scale and
  \item
    $\tau$ denotes the temporal scale.
  \end{itemize}
Figure~\ref{fig-veladapt-space-time-kernel-1+1-D} shows examples of
non-separable spatio-temporal receptive fields measured by cell
recordings in V1 with corresponding velocity-adapted spatio-temporal
receptive fields obtained using the Gaussian scale-space and the
time-causal scale-space; see also \citet{YouLesMey01-SV} and
\citet{YouLes01-SV} for a closely related approach based on Gaussian
spatio-temporal derivatives although using a different type of
parameterization and \citep{Lin97-ICSSTCV} for closely related earlier work.
These scale-space models should be regarded as 
{\em idealized functional and phenomenological models
of receptive fields\/} that predict how computations occur in a visual
system and whose actual realization can then be implemented in different ways
depending on available hardware or wetware.

Work has also been performed on learning receptive field properties 
and visual models from the statistics of natural image data 
\citep{Fie87-JOSA,SchHat96-VisRes,OlsFie96-Nature,RaoBal98-CompNeurSyst,SimOls01-AnnRevNeurSci,Wil08-AnnRevPsychol}
and been shown to lead to the formation of similar receptive fields as
found in biological vision.
The proposed theoretical model on the other hand makes it possible to
determine such receptive fields from theoretical first principles that
reflect symmetry properties of the environment and thus
without need for any explicit training stage or selection of
representative image data.
This normative approach can therefore be seen as describing the
solution that an idealized learning based system may converge to, if
exposed to a sufficiently large and representative set of natural image data.

An interesting observation that can be made from the similarities 
between the receptive field families derived by necessity from the
assumptions and receptive profiles found by cell recordings in
biological vision, is that receptive fields in the
retina, LGN and V1 of higher mammals are very close to {\em ideal\/} in 
view of the stated structural requirements/symmetry properties
\citep{Lin11-RecFields}.
In this sense, biological vision can be seen as having adapted very
well to the transformation properties of the outside world and the
transformations that occur when a three-dimensional world is projected
to a two-dimensional image domain.

\section{Mechanisms for obtaining true geometric invariances}

An important property of the above mentioned families of spatial and
spatio-temporal receptive
fields is that they obey
basic {\em covariance properties\/} under
  \begin{itemize}
  \item
     {\em rescalings\/} of the spatial and temporal dimensions,
  \item
     {\em affine transformations\/} of the spatial domain and
  \item
     {\em Galilean transformations\/} of space-time;
 \end{itemize}
see \citet[section~5.1.2, page~56]{Lin10-JMIV} for more precise
statements and explicit equations.
These properties do in turn allow the vision system to handle:
\begin{itemize}
\item
    image data acquired with different spatial and 
    temporal {\em sampling rates\/}, including
    image data that are sampled with different {\em spatial resolution\/} on a
    foveated sensor with decreasing sampling rate towards the
    periphery and spatio-temporal events
    that occur at {\em different speed\/} (fast {\em vs.\/} slow),
   \item
     image structures of different spatial and/or temporal {\em extent\/}, 
     including objects of different {\em size\/} in the world and
     events with longer or shorter {\em duration\/} over time,
  \item
     objects at different {\em distances\/} from the camera,
  \item
     the linear component of {\em perspective deformations\/} 
     ({\em e.g.\/} perspective foreshortening) and
  \item
     the linear component of {\em relative motions\/} between objects in the
     world and the observer.
\end{itemize}
In these respects, the presented receptive field models ensure that
visual representations will be {\em well-behaved\/} under 
{\em basic geometric transformations\/} in the image formation process.

This framework can then in turn be used as a basis for defining 
{\em truly invariant representations\/}. In the following, we shall describe
basic approaches for this that have been developed in the area of computer
vision, and have been demonstrated to be powerful mechanisms for achieving
scale invariance, affine invariance and Galilean invariance for
real-world data.
Since these mechanisms are expressed at a functional level of
receptive fields, we propose that corresponding mechanisms can be
applied to neural models and a for providing a mathematically
well-founded framework for explaining invariance properties in
computational models.

\subsection{Scale invariance}
\label{sec-sc-inv}

Given a set of receptive fields that operate over some range of scale, a general
approach for obtaining scale invariance is by performing 
{\em scale selection\/} from local extrema over scale of
{\em scale-normalized derivatives\/} \citep{Lin93-Dis,Lin97-IJCV}
\begin{equation}
  \label{eq-sc-norm-ders-def}
  \partial_{\xi_1} = s^{\gamma/2} \, \partial_{x_1}
  \quad\quad
  \partial_{\xi_2} = s^{\gamma/2} \, \partial_{x_2}
\end{equation}
where $\gamma \in [0, 1]$ is a free parameter that can be adjusted to
the task and in some cases can be chosen as $\gamma = 1$.
Specifically, it can be shown that if a spatial image $f(x)$ has a
local extremum over scale at scale $s_0$ for some position $x_0$ in
image space, then if we define a rescaled image $f'(x')$ by
$f'(x') = f(x)$ where $x' = \alpha \, x_0$ for some scaling factor
$\alpha$, then there will be a corresponding local extremum over scale
in the rescaled image $f'(x')$ at scale $s_0' = \alpha^2 s_0$ and
position $x' = \alpha \, x_0$
\citep[section~13.2.1]{Lin93-Dis} \citep[section~4.1]{Lin97-IJCV}.
In other words, local extrema over scale of scale-normalized
derivatives are preserved under scaling transformations and follow the
scale variations in an appropriate manner.
This property also extends to linear and non-linear combinations of
receptive field responses that correspond to spatial and
spatio-temporal derivatives of the Gaussian spatial and
spatio-temporal scale-space concepts described in
section~\ref{sec-ideal-rec-fields} as well as to the idealized
models of biological receptive fields presented in section~\ref{sec-model-biol-rec-fields}.

\begin{figure}[hbtp]
  \begin{center}

     \begin{tabular}{c}
       \includegraphics[width=0.50\textwidth]{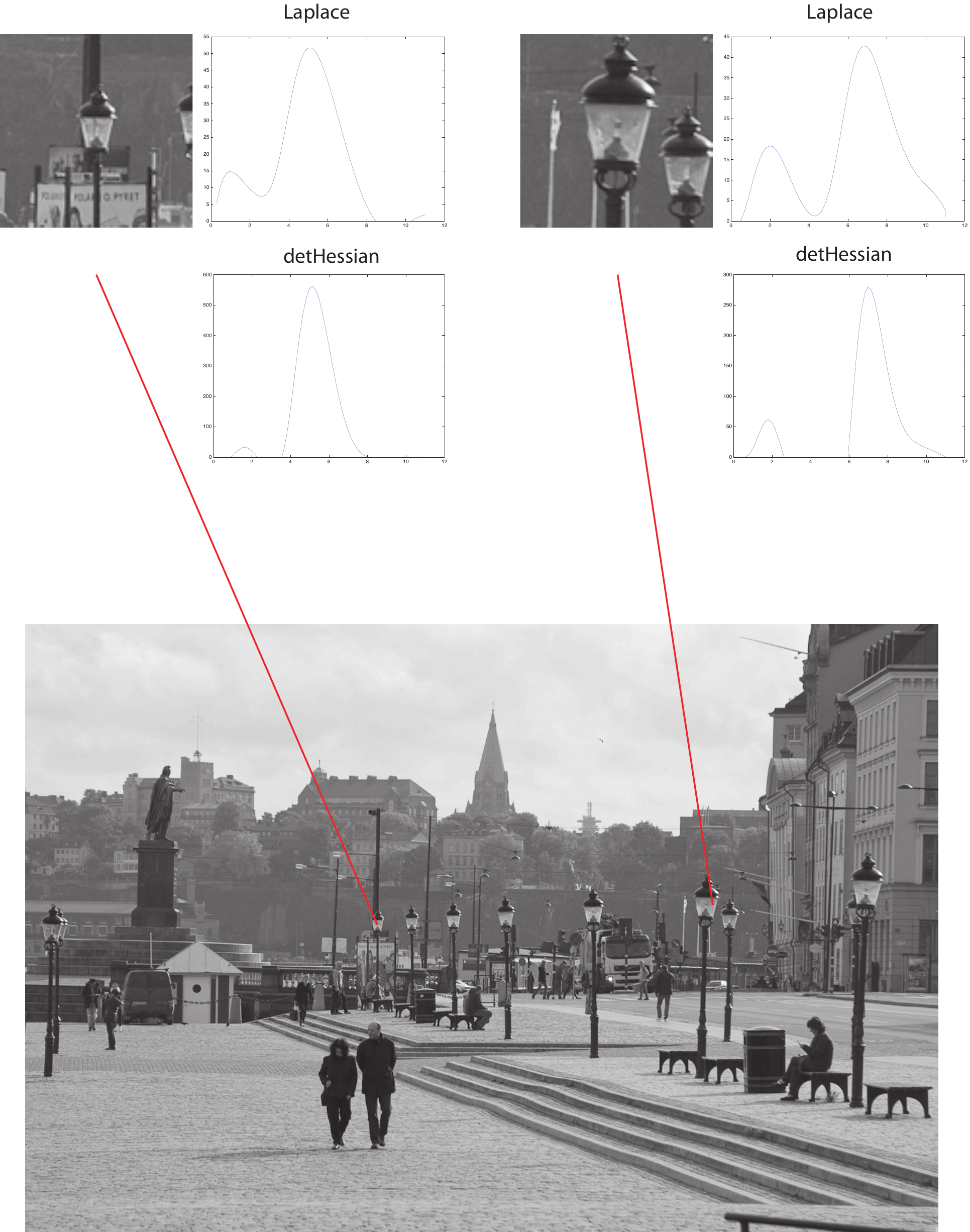} 
     \end{tabular} 

   \end{center}
   \caption{Illustration of how {\em scale selection\/} can be performed from
     receptive field responses by computing scale-normalized Gaussian derivative
     operators at different scales and then detecting local extrema
     over scale. Here, so-called {\em scale-space signatures\/} have been
     computed at the centers of two different lamps at different
     distances to the observer. }
 \label{fig-skeppsbron1-scspsign-laplace-dethess}
 \end{figure}

\begin{figure}[hbtp]
  \begin{center}
     \begin{tabular}{ccccc}
        {\footnotesize\em original window\/}
        & $$
        & {\footnotesize\em scale estimate\/}
        & $$
        & {\footnotesize\em scale normalized\/} \\
        \\
    \includegraphics[width=0.10\textwidth]{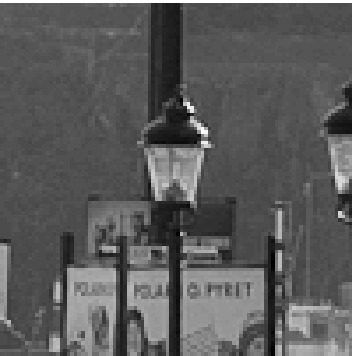}
       & $\quad \quad$
       & \includegraphics[width=0.125\textwidth]{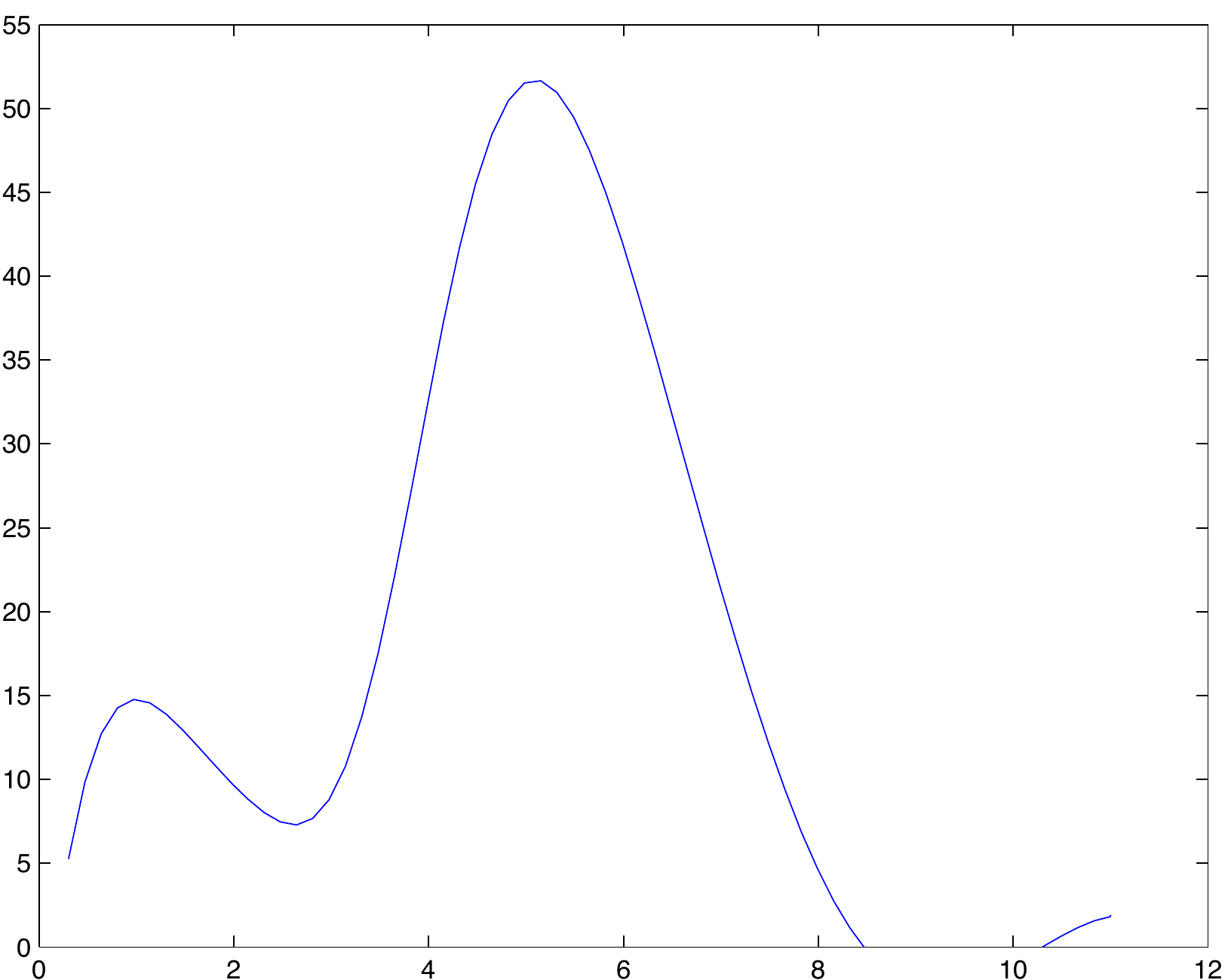}
       & $\quad\quad$
       & \includegraphics[width=0.10\textwidth]{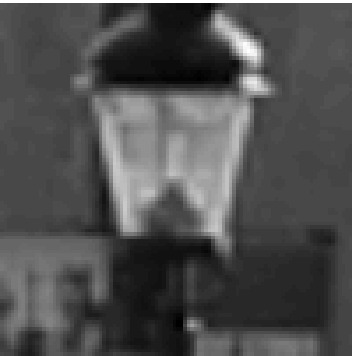}
          \\
      \includegraphics[width=0.10\textwidth]{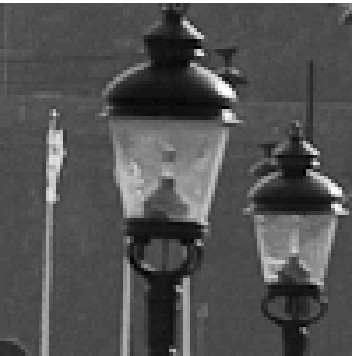}
       & $\quad \quad$
       & \includegraphics[width=0.125\textwidth]{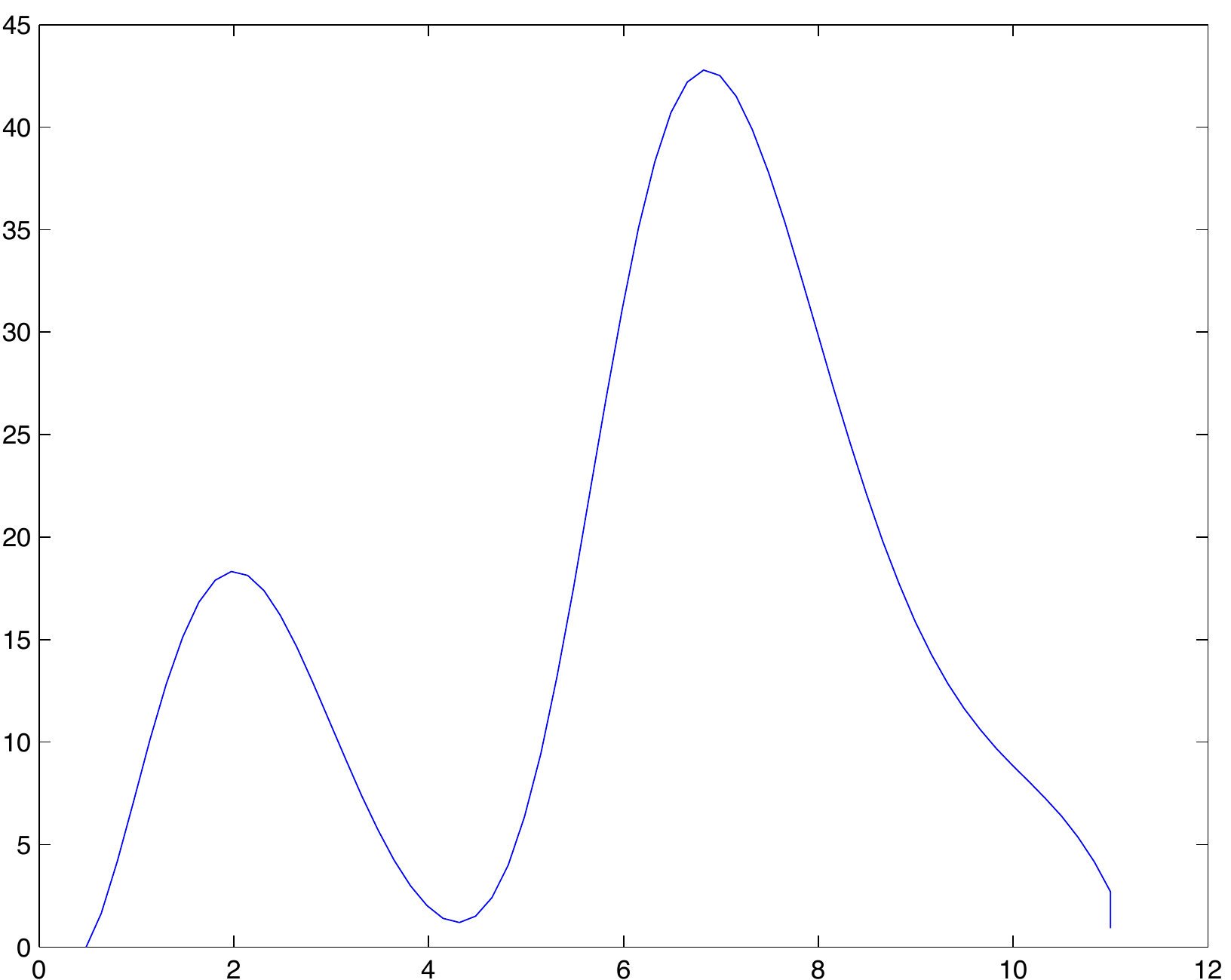}
       & $\quad\quad$
       & \includegraphics[width=0.10\textwidth]{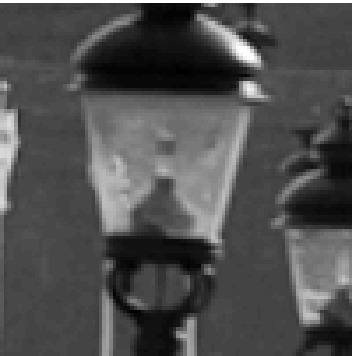}
          \\
    \includegraphics[width=0.10\textwidth]{skeppsbron1-before-scalenorm-small}
       & $\quad \quad$
       & \includegraphics[width=0.125\textwidth]{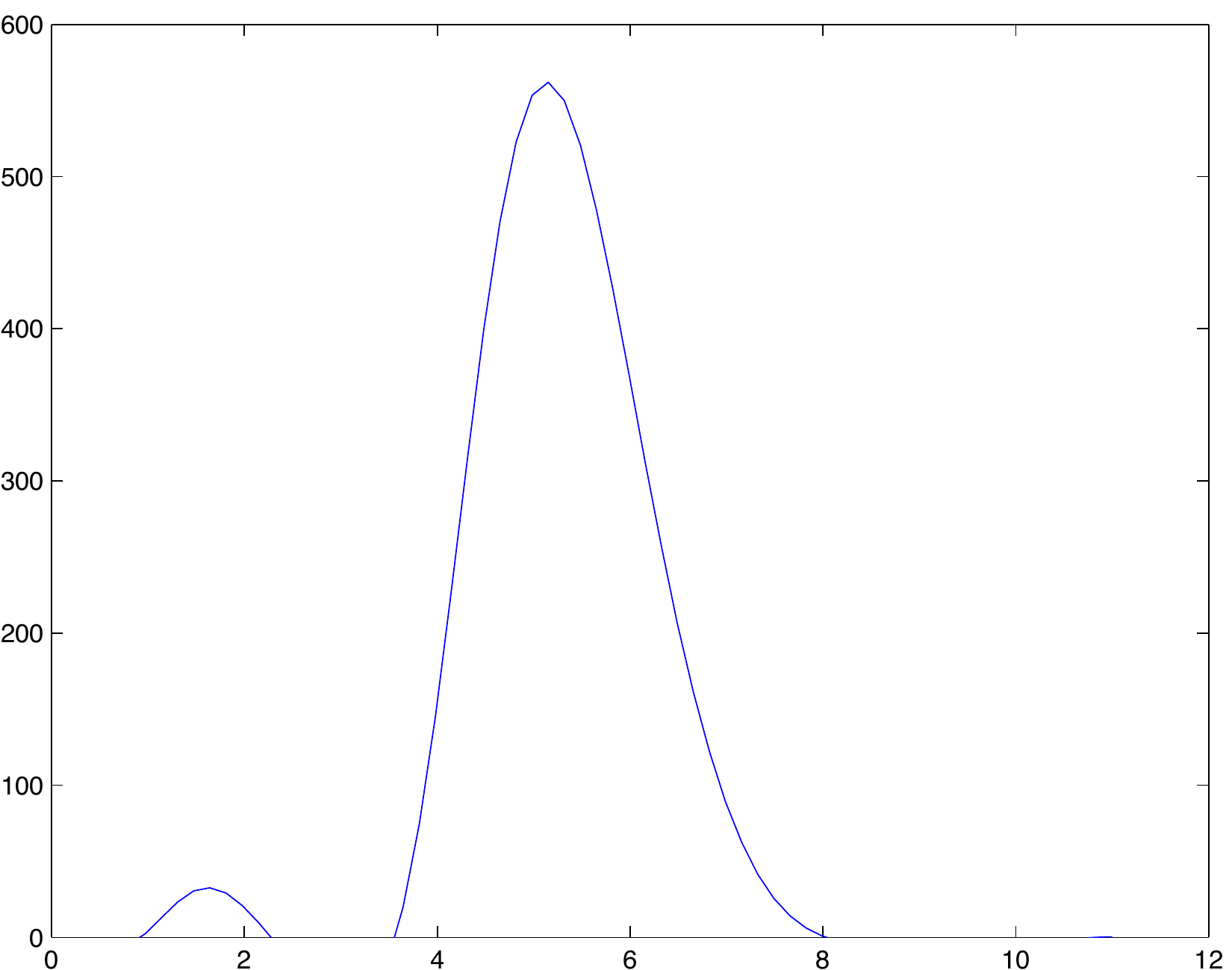}
       & $\quad\quad$
       & \includegraphics[width=0.10\textwidth]{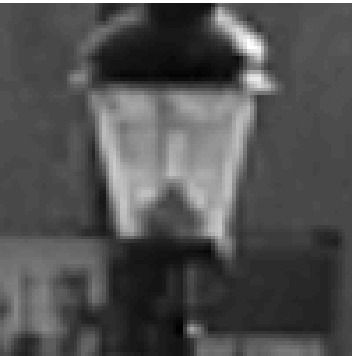}
        \\
     \includegraphics[width=0.10\textwidth]{skeppsbron1-before-scalenorm-large}
       & $\quad \quad$
       & \includegraphics[width=0.125\textwidth]{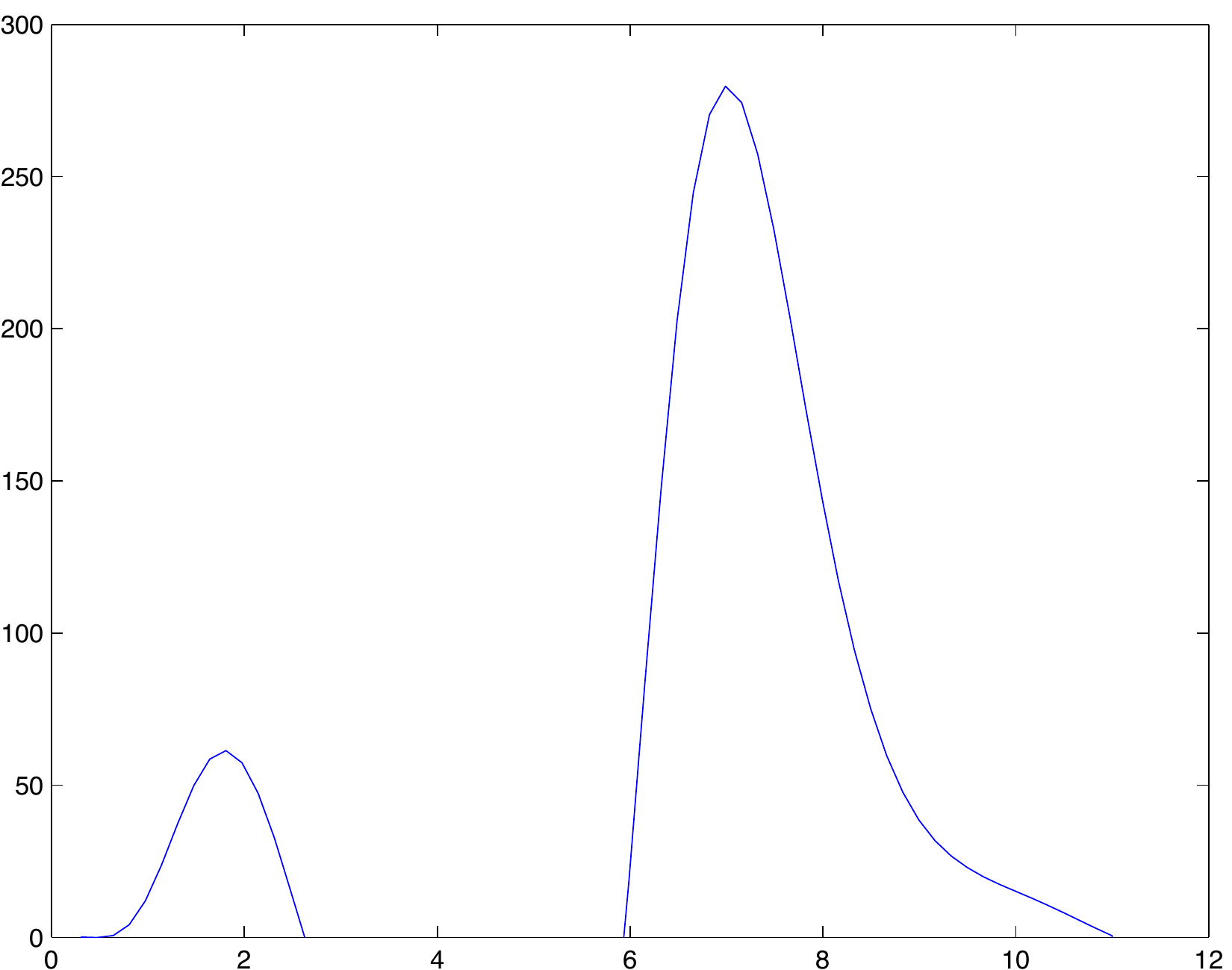}
       & \quad\quad
       & \includegraphics[width=0.10\textwidth]{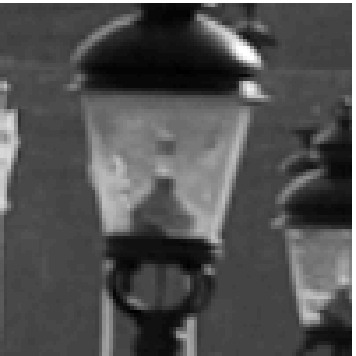}
          \\
    \end{tabular}
  \end{center}
 \caption{Illustration of how {\em scale normalization\/} can be
   performed by rescaling local image structures using scale
   information obtained from a scale selection mechanism.
   Here, the two windows selected in
   figure~\ref{fig-skeppsbron1-scspsign-laplace-dethess} have been
   transformed to a common {\em scale-invariant reference frame\/} by normalizing them
   with respect to the scale levels at which the scale-normalized
   Laplacian and the scale-normalized determinant of the Hessian
   respectively assumed their global extrema over scale. 
   Note the similarities of the resulting scale normalized
   representations, although they correspond to physically different
   objects in the world.}
 \label{fig-scale-norm-windows-skeppsbron-laplace-dethess}
  \vspace{-4mm}
\end{figure}

Figure~\ref{fig-skeppsbron1-scspsign-laplace-dethess} illustrates this idea by performing local scale
selection at two different points in a spatial image from local
extrema over scale of 
the scale-normalized Laplacian $ \nabla_{norm}^2 L$ and 
the scale-normalized determinant of the Hessian $\det {\cal H}_{norm} L$
computed from Gaussian-derivative receptive fields
at different spatial scales
\begin{equation}
  \label{eq-sc-norm-Lapl-detHess}
  \nabla_{norm} L = s (L_{x_1 x_1} + L_{x_2 x_2}),
  \quad\quad
  \det {\cal H}_{norm} L = s^2 (L_{x_1 x_1} L_{x_2 x_2} - L_{x_1 x_2}^2).
\end{equation}
From the graphs in this figure, which show the variation over scale of the scale-normalized Laplacian
$\nabla_{norm}^2 L$ and the
scale-normalized determinant of the Hessian 
$\det {\cal H}_{norm} L$
as function of effective scale $\log s$, it can be seen that the
local extrema over scale are assumed at a finer scale for the
distant object and at a coarser scale for the nearby object.
The ratio between these scale values measured in units of
the standard deviation $\sigma = \sqrt{s}$ of the underlying Gaussian
kernels corresponds to the ratio between the sizes of the
projections between of the two lamps in the image domain
and reflects the ratio between the
distances between these objects and the observer.

Computing image descriptors at scale levels obtained from a scale
selection step based on local extrema over scale of scale-normalized
receptive field responses or equivalently computing image descriptors
from local image patches that have been scale normalized with
respect to such size estimates constitutes a very general approach for
obtaining true {\em scale invariance\/} and has been successfully
applied for different tasks in computer vision \citep{Lin99-CVHB,Lin08-EncCompSci}
including scale-invariant tracking and object recognition
\citep{BL97-CVIU,Low04-IJCV,BayEssTuyGoo08-CVIU} and estimation of time
to collision from temporal size variations in the
image domain \citep{LinBre03-ScSp,NegBraCroLau08-ExpRob}.

Figure~\ref{fig-scale-norm-windows-skeppsbron-laplace-dethess}
illustrates an application of the latter scale normalization approach
applied to the two windows in
figure~\ref{fig-skeppsbron1-scspsign-laplace-dethess}, 
by first detecting local extrema over scale of the scale-normalized Laplacian
$\nabla_{norm}^2 L$ and the
scale-normalized determinant of the Hessian 
$\det {\cal H}_{norm} L$
and then using these scale values for rescaling the two windows to
a common reference frame.
In theory any image measurement derived from the common reference
frame will be truly scale invariant.
Scale selection performed in this way does hence constitute a very general principle for
achieving scale invariance for image measurements in terms of
receptive fields.

It should be emphasized, however, that there is in principle no need
for carrying out the image warping in practice as it has been done in 
figure~\ref{fig-scale-norm-windows-skeppsbron-laplace-dethess} for
the purpose of illustration.
On a neural architecture it may be more efficient to consider a
routing mechanism \citep{OlsAndEss93-JNeurSci,Wis04-ProblSystNeurSci} 
that operates on image representations at different
scales and selects visual representations from the scales at which
image features assume their extremum responses over scale.
In this respect, the resulting model will be qualitatively rather
similar to the approach by \citet{RiePog99-Nature}, where a SoftMax
operation (a soft winner-take-all mechanism) is applied for computing receptive field representations at
successively higher layers in a hierarchical architecture.
Specifically, the notion of scale-normalized derivatives
according to (\ref{eq-sc-norm-ders-def}) determines how
the receptive field responses as modelled by Gaussian derivatives
should be normalized between different scale levels in such a model.%
\footnote{In practice, the scale normalization in 
                equation~(\protect\ref{eq-sc-norm-Lapl-detHess})
                with $\gamma = 1$ corresponds to normalizing the
                underlying Gaussian derivative receptive fields
                (\protect\ref{eq-gauss-ders-2D})
                to constant $L_1$-norm over scale, whereas other
                values of $\gamma \neq 1$ correspond to other
                $L_p$-norms being constant over scale
                \protect\citep[section~9.1, pages~107--108]{Lin97-IJCV}.}
Due to the scale covariant nature of the underlying receptive fields,
it follows that the visual representations that are routed
forward by the
maximum selection mechanism will be truly scale invariant.
Concerning the possible biological implementation of such a maximum
operation, \citet{GawMar02-JNeurPhys} have shown that
there are neurons in area V4 of monkey that respond to two
simultaneously presented stimuli that are well predicted by the
maximum of the response to each stimulus presented separately.

\subsection{Affine invariance}

Given a set of spatial receptive fields as generated from affine
Gaussian kernels (\ref{eq-aff-gauss-kernel}) with their directional
derivatives (\ref{eq-dir-der-order-N}) for different spatial extents
and orientations as specified by different covariance matrices 
(\ref{eq-aff-cov-mat-2D}), the vision system will be faced with the
task of interpreting the output from the corresponding family of receptive field
responses. For example, if we assume that the vision system observes a
local surface patch in the world, one may ask if some specific
selection of filter parameters would be particularly suitable for
interpreting the data in any given situation. Specifically, if the vision system observes the
same surface patch from two different viewing directions, it would be
valuable if the vision system could maintain a stable perception of
the surface patch although it will be deformed in different ways in the two
perspective projections onto the different image planes.

One way of selecting filter responses from such a family of affine
receptive fields is by using image measurements in terms of the
{\em second-moment matrix\/} (structure tensor)
\begin{equation}
  \label{def-mu}
  \mu(x;\; s_{der}, s_{int}) 
  = \int_{u \in \bbbr^2}
         (\nabla L)(u;\; s_{der}) \, (\nabla L)(u;\; s_{der})^T \,
     g(x-u;\; s_{int}) \, du
\end{equation}
where $s_{der}$ is a local scale parameter describing the scale at which
spatial derivatives are computed and $s_{int}$ is a second integration
scale parameter over which local statistics of spatial derivatives is
accumulated \citep{Lin93-Dis,LG96-IVC}.
These statistics correspond to weighted averages of the non-linear
combinations of partial derivatives $L_{x_1}^2$, $L_{x_1} L_{x_2}$ and
$L_{x_2}^2$ using a Gaussian function as weight, and could on a
biological architecture be performed in a visual area that is based on
input from V1, for example in V2.

A useful property of the second-moment matrix is that it transforms in
a suitable way under affine transformations, as will be described next. 
If we consider an affine extension of the second-moment matrix by replacing the scalar scale parameters
$s_{der}$ and $s_{int}$ in (\ref{def-mu}) by corresponding covariance
matrices $\Sigma_{der}$ and $\Sigma_{int}$
\begin{equation}
  \label{def-mu-aff}
  \mu(x;\; \Sigma_{der}, \Sigma_{int}) 
  = \int_{u \in \bbbr^2}
         (\nabla L)(u;\; \Sigma_{der}) \, (\nabla L)(u;\; \Sigma_{der})^T \,
     g(x-u;\; \Sigma_{int}) \, du
\end{equation}
and consider two images $f(x)$ and $f'(x')$ that are related by an
affine transformation $x' = A \, x$ such that $f'(A x) = f(x)$,
then the corresponding affine second-moment matrices will be related
according to 
\begin{equation}
  \label{eq-transf-prop-non-uni-sec-mom-mat-lin-trans}
  \mu'(A \, x;\; A \, \Sigma_{der} \, A^T, A \, \Sigma_{int} \, A^T) 
  = A^{-T} \, \mu(x;\; \Sigma_{der}, \Sigma_{int}) \, A^{-1}.
\end{equation}
Specifically, if we can determine covariance matrices $\Sigma_{der}$ and
$\Sigma_{int}$ such that 
$\mu(x;\; \Sigma_{der}, \Sigma_{int}) = c_1 \, \Sigma_{der}^{-1} = c_2 \, \Sigma_{int}^{-1}$ 
for some constants $c_1$ and $c_2$, we obtain a {\em fixed-point\/}
that will be {\em preserved under affine transformations\/}
\citep{Lin93-Dis,LG96-IVC}.
This property can be used for signalling if the image measurements
that have been performed for a particular setting of filter parameters in a
family of affine Gaussian receptive fields satisfy the fixed-point
requirement. If so, they can be used for defining an 
{\em affine invariant reference frame\/}%
\footnote{If the local image pattern is weakly isotropic in the sense
  that the second-moment computed in the tangent plane of the surface
  is proportional to the unit matrix $\mu_{surf} = c \, I$ for some
  constant $c$ \citep{GL94-IJCV}, then the foreshortening caused by the perspective
  foreshortening will be compensated for by the affine transformation given by
  (\ref{eq-transf-prop-non-uni-sec-mom-mat-lin-trans}). 
  For non-isotropic image patterns with 
  $\mu_{surf} \neq c \, I$ this interpretation no longer holds,
  but the affine transformed surface pattern will still be affine invariant.}
by transforming the local
image patch with a linear transformation proportional to $A =
\mu^{1/2}$.

It should be noted, however, that the affine transformation $A$ is not
uniquely determined by the fixed-point requirement (\ref{def-mu-aff}),
which only determines two of the four parameters, corresponding to
amount and direction of perspective foreshortening of a local surface
pattern, in other words the viewing direction in relation to an
object centered coordinate system.
The two remaining degrees of freedom correspond to 
(i)~an overall scaling factor corresponding the viewing distance, 
which can be determined by scale selection as described in 
section~\ref{sec-sc-inv}, and 
(ii)~a free rotation angle, corresponding to the selection of a
representative direction in the image plane.
If the vertical direction is preserved under the perspective
transformation, we may therefore not need to determine it 
or just adjust it from an initial estimate.

\begin{figure}[hbt]
  \vspace{-1mm}
  \begin{center}
    \begin{tabular}{ccc}
      {\footnotesize\em oblique view of an object}
      & {\footnotesize\em affine invariant reference frame\/} 
      & {\footnotesize\em invariant receptive field response\/} \\
      \hspace{-2mm} \includegraphics[scale=0.393]{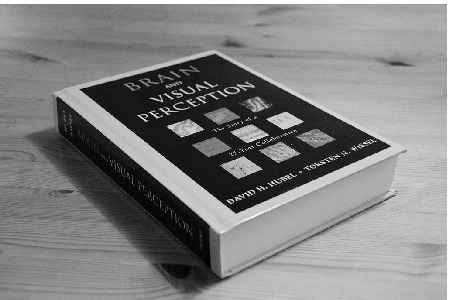} \hspace{-2mm} &
      \hspace{-2mm} \includegraphics[scale=0.33]{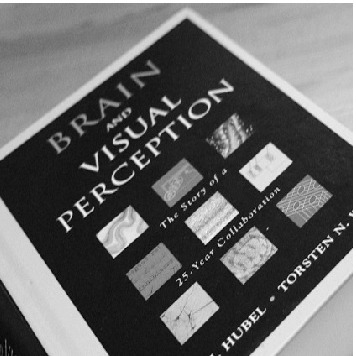} \hspace{-2mm} &
      \hspace{-2mm} \includegraphics[scale=0.33]{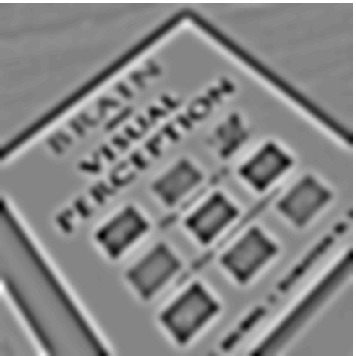} \hspace{-2mm} \\
    \end{tabular} 
  \end{center}
  \caption{Illustration of how {\em affine invariance\/} can be achieved by
     normalization to an {\em affine invariant reference frame\/} determined
     from a second-moment matrix. 
     (left) A grey-level image with an oblique view of a book cover. 
     (middle) The result of affine normalization of a central image
     patch using a series of affine transformations proportional to 
     $A = \mu^{1/2}$ until the an affine invariant fixed-point of
     (\protect\ref{eq-transf-prop-non-uni-sec-mom-mat-lin-trans}) has been reached.
    (right) An example of computing receptive field responses,
    here the Laplacian $\nabla^2 L$, in the affine normalized
     reference frame. Receptive field responses computed in this
     reference frame will be affine invariant up to
     an undetermined scaling factor and an undetermined rotation angle
   and therefore invariant with respect to the viewing direction of
   the object in
   relation to an object centered coordinate system.}
  \label{fig-book-affineadapt}
\end{figure}

Figure~\ref{fig-book-affineadapt} illustrates an application of this
idea to an image with an oblique view of a book cover.
Here, a second-moment matrix $\mu$ has been computed over 
a central window in the original image shown in the left figure.
Then, an affine transformation matrix $A = \mu^{1/2}$ has been used
for warping the central window to a reference frame, and this
so-called {\em affine shape-adaptation\/} process has been repeated until
the second-moment matrix in the reference frame is sufficiently close
to proportional to the unit matrix $\mu' \approx c \, I$.
The middle figure shows the result of warping a central window in the
original image to the resulting affine normalized reference frame.
Due to the affine invariant property of the fixed point
(\ref{eq-transf-prop-non-uni-sec-mom-mat-lin-trans}), 
any receptive field response computed in this reference frame will be
affine invariant up to an undetermined scaling factor and a free
rotation angle.
Hence, this method provides a way of normalizing receptive field
responses with respect to image transformations outside the similarity
group that correspond to variations in the viewing direction relative
to the object.

Again it should, however, be emphasized that it is in principle not
needed to perform the actual image transformation in reality 
to achieve the affine invariant property. 
On a neural architecture that computes a family of affine receptive
fields with different orientations and spatial extents in parallel, one can again
consider a routing mechanism that selects the receptive field
responses from those receptive fields whose measurements of
second-moment matrices are in best agreement with the underlying
covariance matrices in relation to the fixed-point property.
Then, up to a known transformation whose parameters can be computed from the
corresponding second-moment matrix, these routed receptive field
responses will also be affine invariant.

In the area of computer vision, this idea of affine shape adaptation has been used for defining
affine invariant image descriptors with successful applications to image matching, recognition and
estimation of cues to surface shape
\citep{LG96-IVC,Bau00-CVPR,MikSch04-IJCV,TuyGoo04-IJCV,LazSchPon05-PAMI,RotLazSchPon06-IJCV}.

On a neural architecture, one can also conceive that a neuron or a
group of neurons that are adapted to a particular shape of the
covariance matrix corresponding to an orientation in space could
determine if the local image measurements that have been performed for
this particular orientation in space are in agreement with the
fixed-point requirement
(\ref{eq-transf-prop-non-uni-sec-mom-mat-lin-trans}).
If so, the neuron(s) could respond with a high activity if the local image
measurements agree with the filter parameters to which the receptive
fields are tuned and with a low activity otherwise.
Hence, this framework allows for the formulation of affine invariant
receptive field responses, to support view-invariant recognition at
the level of groups of oriented receptive fields over a set of
different covariance matrices $\Sigma_k$.

\subsection{Galilean invariance}

\begin{figure}[!p]

  \begin{center}
    \begin{tabular}{c}
       \includegraphics[width=0.32\textwidth]{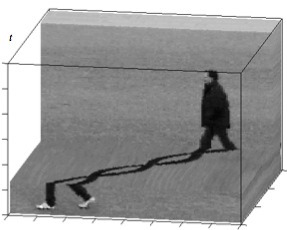}
    \end{tabular}

    \begin{tabular}{ccc}
        {\footnotesize\em stabilized camera\/}
          &
          & {\footnotesize\em stationary camera\/} \\
        \includegraphics[width=0.25\textwidth]{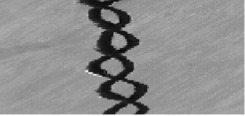}
        & {\footnotesize\multirow{2}{35mm}[12mm]{\em gait pattern in\\ $(x, t)$-plane\/}}
        & \includegraphics[width=0.25\textwidth]{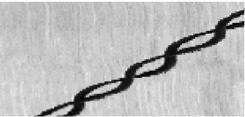} \\
        \includegraphics[width=0.25\textwidth]{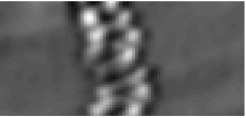}
        & {\footnotesize\multirow{2}{35mm}[12mm]{\em space-time separable receptive field responses\/}}
        & \includegraphics[width=0.25\textwidth]{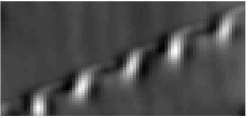}  \\
        $\,$
        & {\footnotesize\multirow{2}{35mm}[12mm]{\em warped to\\ stabilized frame\/}}
        & \includegraphics[width=0.25\textwidth]{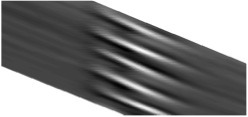} \\
        \includegraphics[width=0.25\textwidth]{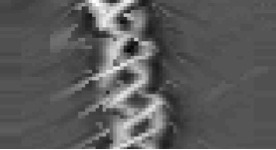}
        & {\footnotesize\multirow{2}{35mm}[12mm]{\em locally velocity-adapted\\ receptive field responses\/}}
        & \includegraphics[width=0.25\textwidth]{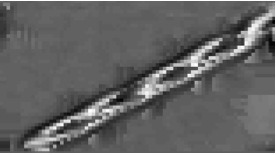} \\
        $\,$
        & {\footnotesize\multirow{2}{35mm}[12mm]{\em warped to\\ stabilized frame\/}}
        & \includegraphics[width=0.25\textwidth]{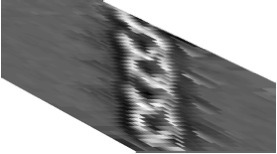} \\
    \end{tabular}
  \end{center}

  \caption{Illustration of how receptive field responses may be
    affected by unknown relative motions
    between objects in the world and the observer and of how this
    effect can be handled by {\em local velocity adaptation\/}.
    The first row shows space-time traces of a walking person taken with
    (left column) a stabilized camera with the viewing direction
    following the motion of the person and
    (right column) a stationary camera with a fixed viewing direction.
    The second row shows Laplacian receptive field responses computed in the two
    domains from space-time separable receptive fields without
    velocity adaptation.
    In the third row, these receptive field responses from the
    stationary camera have been space-time warped to the reference
    frame of the stabilized camera.
    As can be seen from the data, the receptive field responses are
    quite different in the two domains, which implies problems if one would try to match
    them. Hence, spatio-temporal recognition based on 
    space-time separable receptive fields only can be a rather
    difficult problem.
    In the fourth row, the receptive field responses have instead been
    computed with {\em local velocity adaptation\/}, by
    computing extremum responses of the Laplacian receptive field
    responses over different image velocities for each point in space-time.
    In the fifth row, the velocity-adapted receptive responses from the
    stationary camera have been space-time warped to the reference frame of the stabilized camera.
    As can be seen from a comparison with the corresponding result
    obtained for the non-adapted receptive field responses in 
    the third row,
    the use of velocity adaptation implies a better stability of
    receptive field responses under unknown relative motions between
    objects in the world and the observer.
    (Adapted from \protect\citep{LapLin03-IVC}.)}
    \label{fig-st-traces-rec-fields}
\end{figure}

Given a family of spatio-temporal receptive field that are adapted to
motions of different image
velocities $v$ and given an object that moves with some unknown image
velocity $u$ in relation to the viewing direction,
the vision system also faces the problem of how to interpret the
output from the family of receptive fields.
Figure~\ref{fig-st-traces-rec-fields}
shows an illustration of how receptive field responses may be affected
by relative motions between objects in the world and the observer.

If we would know the image velocity $u$ of the object beforehand, it could of
course be preferable to select receptive field responses from the
receptive fields that are adapted to precisely this image 
velocity $v = u$.
A priori, we cannot, however, assume such knowledge,
since one of the basic tasks in relation to object recognition may be to
determine the image velocity of an unknown moving object.
There are also classes of composed spatio-temporal events consisting of
different image velocities at different positions $x$ and time
moments $t$ in space-time, for which it may not be trivial how a representative 
image velocity could be defined for the spatio-temporal event as a
whole.
Hence, this problem warrants a principled treatment.

Given spatio-temporal image data $f(x, t) = f(p)$ with a position in
space-time denoted by $p = (x, t)$, let us define a 
Gaussian spatio-temporal scale-space representation $L$ of $f$
by convolution with a Gaussian spatio-temporal kernel 
$g(\cdot;\; \Sigma_{der}, \delta_{der})$ with spatio-temporal
covariance matrix $\Sigma_{der}$ of the form (\ref{eq-spat-temp-cov-matrix-2+1-D-space-time})
and with time delay $\delta_{der}$.
With a {\em spatio-temporal second-moment matrix\/} $\mu$ over
2+1-D space-time defined according to \citep[equation~(191), page~73]{Lin10-JMIV}
\begin{align} 
  \begin{split}
    \label{eq-def-mu-again}
    \mu(p;\; \Sigma_1, \Sigma_2, \delta_1+\delta_2) 
    & = \int_{q \in \bbbr^{(2+1)}}
        (\nabla L(q;\; \Sigma_1, \delta_1)) (\nabla L(q;\; \Sigma_1, \delta_1))^T \, 
          g(p - q;\; \Sigma_2, \delta_2) \, dq,
  \end{split} 
\end{align}
where $g(p - q;\; \Sigma_2, \delta_2)$ denotes a second-stage Gaussian
smoothing with covariance matrix $\Sigma_2$ and time delay $\delta_2$
over space-time, it is indeed possible to perform such velocity selection.
Consider two Galilean-related spatio-temporal image data sets 
$f'(p') = f(p)$ that are related by a relative image velocity 
$u - v$ such that $p' = G_{u-v} \, p$ for a Galilean transformation
matrix $G_{u-v}$ according to (\ref{eq-def-gal-transf}).
Then, it can be shown that the
corresponding spatio-temporal covariance matrices are related
according to \citep[equation~(193), page~73]{Lin10-JMIV}
\begin{equation}
  \label{eq-mubiss-from-mu}
  \mu' = G^{-T}_{u-v} \, \mu \, G^{-1}_{u-v}.
\end{equation}
Let us introduce the notion of 
{\em Galilean diagonalization\/}, which corresponds to 
finding the unique Galilean transformation that
 transforms the spatio-temporal second-moment matrix
 to block diagonal form with all mixed purely spatio-temporal
 components being zero 
$\mu'_{x_1 t} = \mu'_{x_2 t} = 0$ \citep{LinAkbLap04-ICPR}
\begin{equation}
  \label{eq-def-gal-diag}
  \mu' =
  \left(
    \begin{array}{ccc}
      \mu_{x_1 x_1}' & \mu_{x_1 x_2}' & 0 \\
      \mu_{x_1 x_2}' & \mu_{x_2 x_2}' &  0 \\
      0                 & 0                     & \mu_{tt}'  \\
   \end{array}
  \right).
\end{equation}
Such a block diagonalization can be obtained if the velocity vector
$u$ satisfies
\begin{equation}
  \label{eq-def-gal-diag-solve-u}
 \left(
    \begin{array}{cc}
      \mu_{x_1 x_1}' & \mu_{x_1 x_2}' \\
      \mu_{x_1 x_2}' & \mu_{x_2 x_2}' \\
 \end{array}
  \right)
  \left(
    \begin{array}{c}
      u_1 \\
      u_2 \\
  \end{array}
  \right)
  =
  -
  \left(
    \begin{array}{c}
      \mu_{x_1 t} \\
      \mu_{x_2 t} \\
  \end{array}
  \right)
\end{equation}
with the solution
\begin{equation}
  \label{eq-vel-est-gal-diag}
  u = - \{ \mu_{xx} \}^{-1} \{ \mu_{xt} \}
\end{equation}
i.e., structurally similar equations as are used for computing optic flow
according to the method by \citep{LukKan81-IU}.
It can then be shown that
the property of Galilean block diagonalization is preserved under Galilean
transformations \citep[appendix~C.4, pages 73--74]{Lin10-JMIV}.
Specifically, the velocity vector associated with the Galilean 
transformation, that brings a second-moment matrix into block diagonal
form, is additive under superimposed Galilean transformations.
This is a very general approach for normalizing local spatio-temporal image
patterns, which also applies to spatio-temporal patterns that cannot
be modelled by a Galilean transformation of an otherwise temporally
stationary spatial pattern.
Specifically, spatio-temporal receptive field responses that can be
expressed with respect to such a {\em spatio-temporal reference frame\/}
will be Galilean invariant.

These ideas have been applied in computer vision for performing
spatio-temporal recognition under unknown relative motions between the
spatio-temporal events and the observer
\citep{LapLin03-IVC,LapCapSchLin07-CVIU}.
Notably the approach in \citep{LapLin03-IVC} is based on a set of
spatio-temporal receptive fields over which simultaneous selection of
image velocities and spatio-temporal scales is performed.

Again, it is not necessary to carry out the spatio-temporal
normalization in practice to achieve Galilean invariance.
On a neural architecture based on a family of spatio-temporal receptive
fields that operate over some set of image velocities in parallel,
one may consider a routing mechanism that selects receptive field
responses by judging the degree of agreement with the criterion of
Galilean diagonalization (\ref{eq-def-gal-diag}) and then giving
priority to the responses that are most consistent with this criterion.
Notably, such a computational mechanism will have the ability to
respond to different motions at different spatial and temporal scales
and may therefore have the ability to handle transparent motion.

Please, note that all information that is needed for computing the
spatio-temporal second-moment matrix and the Galilean diagonalization
are spatio-temporal averages of the non-linear combinations 
$L_{x_1}^2$, $L_{x_1} L_{x_2}$, $L_{x_2}^2$, 
$L_{x_1} L_t$, $L_{x_2} L_t$ and $L_t^2$
of first-order spatio-temporal derivatives and can hence be computed
from spatio-temporal receptive fields.
On a biological architecture, the corresponding information could
therefore be computed from the output of V1 neurons in combination 
with an additional layer of spatio-temporal smoothing.
Thus, similar type of information could in principle be computed by a
visual motion area with direct access to the output from V1, such as V5/MT.

On a neural architecture, one can also conceive that a neuron or a
group of neurons that are adapted to a particular image velocity could
determine if the local spatio-temporal image measurements that have been performed for
this particular image velocity in space-time are in agreement with the
fixed-point requirement (\ref{eq-def-gal-diag}) of
Galilean diagonalization.
If so, the neuron(s) could respond with a high activity if the local
measurements agree with the filter parameters to which the receptive
fields are tuned and with a low activity otherwise.
Hence, this framework allows for the formulation of Galilean invariant
neurons, to support invariant recognition of visual
objects under unknown relative motions between the object and the
observer, provided that this invariance property is formulated at
the level of groups of oriented receptive fields over a set of
image velocities $v_k$.

\section{Invariance property under illumination variations}
\label{sec-inv-illum-var}

In the treatment so far, we have described how image measurements in
terms of receptive fields are related to the geometry of space and
space-time, under the assumption that the actual image intensities
from which the receptive field responses are to be computed have been given beforehand.
One may, however, consider alternative ways of parameterizing the
intensity domain by monotonous intensity transformations that preserve
the ordering between the image intensities,
and in this respect would contain essentially equivalent information.

Given the huge range of luminosity variations under natural imaging
conditions (corresponding to a range of the order of $10^{10}$ between
the darkest and brightest cases for human vision), it is natural to
represent the image luminosities on a 
{\em logarithmic luminosity scale\/}
\begin{equation}
  \label{eq-log-par-intensity}
  \begin{array}{ll}
  f(x) \sim \log I(x)       & \mbox{(time-independent images)}, \\
  f(x, t) \sim \log I(x, t) & \mbox{(spatio-temporal image data)}.
  \end{array}
\end{equation}

\noindent
Specifically, receptive field responses that are computed from such a
logarithmic parameterization of the image luminosities can be {\em interpreted physically\/} as a
superposition of relative variations of surface structure and
illumination variations.
Given a (i)~perspective camera model extended with 
(ii)~a thin circular lens for gathering incoming light from different
directions and 
(iii)~a Lambertian illumination model extended with 
(iv)~a spatially varying albedo factor for modelling the light that is
reflects from surface patterns in the world, 
it can be shown \citep{Lin11-RecFields} that a spatial receptive field response
\begin{equation}
  L_{x^{\alpha}}(\cdot;\; s) 
  = L_{x_1^{\alpha_1}  x_2^{\alpha_2}}(\cdot, \cdot;\; s)
  = \partial_{x^{\alpha}} \, {\cal T}_s \, f
\end{equation}
of the image data $f$, where ${\cal T}_s$ represents the spatial
smoothing operator (here corresponding to a two-dimensional
Gaussian kernel (\ref{eq-2D-gauss})), can be expressed as
\begin{equation}
  \label{eq-illum-model-spat-rec-field}
  L_{x^{\alpha}}
  = \partial_{x^{\alpha}} \, {\cal T}_s \,
   \left(
     \log \rho(x) + \log i(x) 
     + \log C_{cam}(\tilde{f}) 
     + V(x)
  \right)
\end{equation}
where 
\begin{itemize}
\item[(i)]
  $\rho(x)$ is a spatially dependent {\em albedo factor\/} that reflects
  {\em properties of surfaces of objects\/} in the environment with the implicit
  understanding that this entity may in general refer to points on different
  surfaces in the world depending on the viewing direction 
  and thus the image position $x = (x_1, x_2)$,
\item[(ii)]
  $i(x)$ denotes a spatially dependent {\em illumination field\/} 
  with the implicit understanding that the 
  amount of incoming light on different surfaces may be different for
  different points in the world as mapped to corresponding image
  coordinates $x$,
\item[(iii)]
  $C_{cam}(\tilde{f}) = \frac{\pi}{4} \frac{d}{f}$ represents {\em internal camera parameters\/}
  with the ratio $\tilde{f} = f/d$ referred to as the {\em effective
    $f$-number\/}, where $d$ denotes the diameter of the lens and $f$ the
  focal distance, 
\item[(iv)]
  $V(x) = V(x_1, x_2) = - 2 \log (1 + x_1^2  + x_2^2)$ represents a
  geometric {\em natural vignetting\/} effect corresponding to the factor
  $\log \cos^4(\phi)$ for a planar image plane, with $\phi$ denoting
  the angle between the viewing direction $(x_1, x_2, f)$ and the
  surface normal $(0, 0, 1)$ of the image plane. (This vignetting term will disappear
  for a spherical camera model.)
\end{itemize}

\begin{figure}[hbt]
  \vspace{-1mm}
  \begin{center}
    \begin{tabular}{cc}
      {\footnotesize\em linear luminosity scale\/}
      & {\footnotesize\em logarithmic luminosity scale\/} \\
     \includegraphics[width=0.45\textwidth]{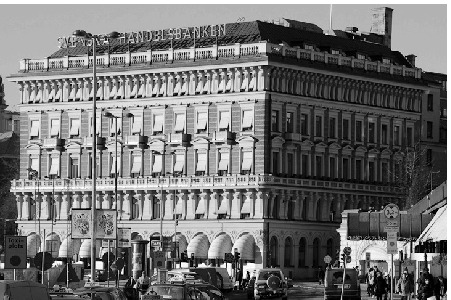} &
      \includegraphics[width=0.45\textwidth]{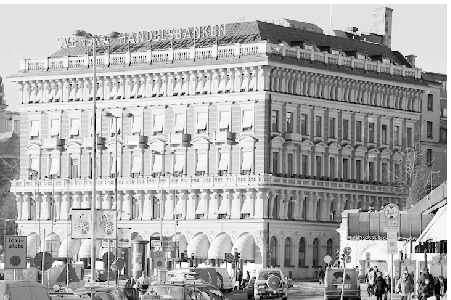} \\
      {\footnotesize\em Laplacian responses from linear luminosities\/} 
      & {\footnotesize\em Laplacian responses from logarithmic luminosities\/} \\
      \includegraphics[width=0.45\textwidth]{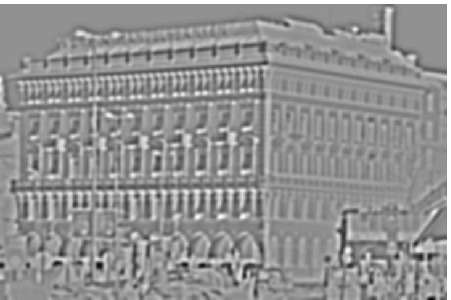} &
      \includegraphics[width=0.45\textwidth]{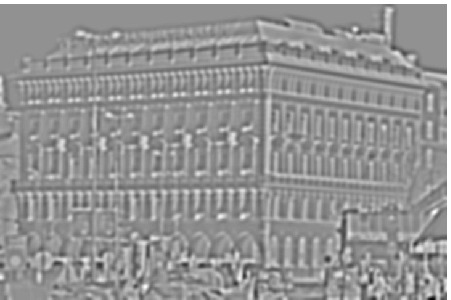} \\
   \end{tabular} 
  \end{center}
  \caption{Illustration of the effect of computing Laplacian receptive
  field responses $\nabla^2 L$ from image intensities defined on 
  (left column) a linear
  intensity scale $f(x) \sim I(x)$ {\em vs.\/} (right column) a logarithmic
  intensity scale $f(x) \sim \log I(x)$ for an image with
  substantial illumination variations.
  As can be seen from the figure, the magnitudes of the Laplacian
  receptive field response are substantially higher in the left sunlit
  part of the house compared to the right part in the shade if the
  Laplacian responses are computed from a linear luminosity scale,
  whereas the difference
  in amplitude is between the left and the right parts of the house becomes
  substantially lower if the receptive field responses are
  computed from a logarithmic intensity scale.
  }
  \label{fig-handbank-logtransf}
\end{figure}

\noindent
From the structure of equation~(\ref{eq-illum-model-spat-rec-field})
we can note that for any non-zero order of differentiation 
$\alpha > 0$, the influence of internal camera parameters in 
$C_{cam}(\tilde{f})$ will disappear because of the spatial
differentiation with respect to $x$, and so will the effects be of any
other multiplicative exposure control mechanism.%
\footnote{For biological vision, such multiplicative exposure control
  mechanisms correspond to adaptations of the luminosity on the
  retina by varying the diameter of the pupil as well as adaptations of the
  light sensitivity of the photoreceptors to the luminosity.}
Furthermore, for any multiplicative illumination variation
$i'(x) = C \, i(x)$, where $C$ is a scalar constant,
the logarithmic luminosity will be transformed as
$\log i'(x) = \log C + \log i(x)$, which implies that
the dependency on $C$ in (\ref{eq-log-par-intensity}) will disappear after spatial differentiation.
Thus, receptive field responses in terms of spatial derivatives are
{\em invariant under multiplicative illumination variations\/}.

Specifically, if the illumination field $i(x)$ is constant over the
support region of the receptive field, the receptive field response
will then up to the variations in the natural vignetting $V(x)$
only respond to the spatial variations of the albedo factor 
$\rho(x)$, {\em i.e.\/}, only to variations in the surface pattern(s)
in the world.
Hence, {\em the receptive field responses will have a direct physical 
interpretation in terms of properties of objects and events in the
environment\/}.
This result can be seen as a theoretical explanation of why recognition
methods based on receptive field responses work so well in the area of
computer vision.
More generally, this result could also be seen as a theoretical
explanation of how the receptive field responses that are computed in
LGN and V1 can constitute the foundation for the visual operations in higher visual areas in
biological vision.

Notably, the vignetting effect $V(x)$ is independent of the image
contents $f$ and could therefore be corrected for given sufficient
knowledge about the camera.
For spatio-temporal receptive fields $\partial_{x^{\alpha} t^{\beta}} L$ 
that involve explicit temporal derivatives with $\beta > 0$,
it will furthermore disappear altogether, 
since the vignetting only depends upon the spatial coordinates.

\section{Summary and conclusions}

We have described how the shapes of receptive field
profiles in the early visual pathway can {\em be constrained\/} from structural
symmetry properties of the environment, which include the requirement 
that the receptive field responses should be sufficiently well-behaved 
(covariant) under basic image transformations.
We have also shown how these covariance properties of receptive fields 
{\em enable true invariance properties\/} of visual processes at the 
systems level,  if combined with max-like operations over the output 
of receptive field families tuned to different filter parameters.

The invariance and covariance properties that we have considered include
(i)~scaling transformations to handle objects and substructures of
different size as well as
objects at different distances from the observer,
(ii)~affine transformations to capture image deformations caused by
the perspective mapping under variations of the viewing direction,
(iii)~Galilean transformation to handle unknown relative motions between objects in the world and
the observer and
(iv)~multiplicative intensity transformations to provide robustness to
slowly varying illumination variations as well as invariance to intensity variations
caused by multiplicative exposure control mechanisms.

These transformations should be interpreted as 
{\em local approximations\/} of the actual image transformations,
which in general can be assumed to be non-linear.
Thus, a sufficient requirement for these invariance or covariance
properties to be hold in practice 
and thus enable robust visual recognition from real-world image data, 
is that these approximations should hold
locally within the support region of a given receptive field.
Therefore, these theoretical results can be extended to more complex
scenes by using different local
approximations for receptive fields
at different spatial or spatio-temporal points.

The presented theory leads to a computational framework for 
defining spatial and spatio-temporal receptive fields from visual data
with the attractive properties that:
(i)~the receptive field profiles can be derived {\em by necessity\/} from
first principles and 
(ii)~it leads to {\em predictions\/} about receptive field profiles in good agreement
with receptive fields found by cell recordings in biological vision.
Specifically, idealized models have been presented for 
space-time separable receptive fields in the retina and LGN 
and for non-separable simple cells in V1.

The modelling performed in this article has been performed at a more
abstract level of computation than used in many other
computational models, and should therefore be applicable to a large
variety of neural models provided that their functional properties can
be described by appropriate diffusion equations. 
These results are therefore very general, since they are based on inherent
properties of the image formation process, and should therefore have
important implications for computational modelling of visual processes 
based on receptive fields.
If one accepts the assumptions underlying the model, these results
should therefore have important implications for computational
neuroscience, since they hold for any computational model whose
functionality is compatible with the assumptions.

Compared to more common approaches of learning receptive field
profiles from natural image statistics, the proposed framework 
makes it possible to derive the shapes
of idealized receptive fields without any need for training data. 
The proposed framework for invariance and covariance properties
also adds explanatory value by showing that the families of receptive profiles 
tuned to different orientations in space and image velocities in
space-time that can be observed in biological vision can be {\em explained\/}
from the requirement that the receptive fields should be covariant
under basic image transformations to enable true invariance
properties.
If the underlying receptive fields would not be covariant, then there
would be a systematic bias in the visual operations, corresponding to
the amount of mismatch between the backprojected receptive fields.


The theory could also be used as a framework for raising questions
concerning invariance properties of biological vision.
As a complement to the fundamental covariance properties, we have
outlined possible mechanisms for how {\em true invariance\/} under scaling
transformations, affine transformations and Galilean transformations
can be obtained already at the level of receptive field responses.
The presented mechanisms are based on two types of major principles;
(i)~by detecting {\em extremum values\/} of appropriately normalized receptive
field responses over variations of the filter parameters or 
(ii)~by {\em normalizing\/} the receptive field responses with respect to a
preferred {\em reference frame\/} that is constructed from criteria that are
invariant under the corresponding image transformations.
These methods have been successfully applied in the area of computer
vision and demonstrate how the covariance properties of the proposed 
receptive field model can be used for defining truly scale invariant, 
affine invariant and Galilean invariant visual operations already at
the level of receptive fields, which can then provide a basis for
computational mechanisms for invariant recognition of visual
objects and events at the systems level.

\begin{figure}[hbt]
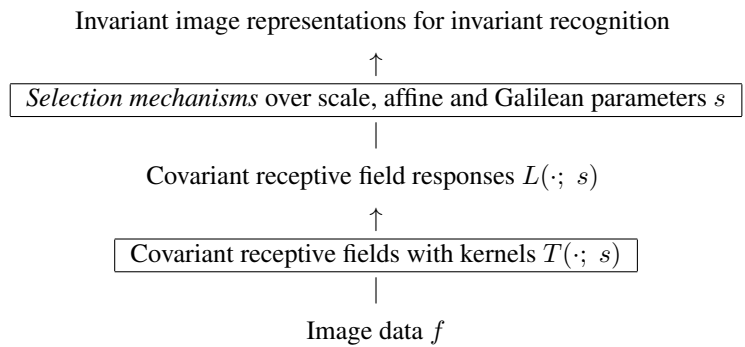

  \begin{center}

  \small

  \vspace{1.4mm}
  \begin{tabular}{c}
    Invariant image representations for invariant recognition\\
  \end{tabular}
  \vspace{1.4mm}

  \begin{tabular}{c}
    $\uparrow$
  \end{tabular}

  \begin{tabular}{|c|}
    \hline
    {\em Selection mechanisms\/} over scale, affine  and
    Galilean parameters $s$ \\
   \hline
  \end{tabular}

  \begin{tabular}{c}
    $\mid$
  \end{tabular}

  \vspace{1.4mm}
  \begin{tabular}{c}
    Covariant receptive field responses $L(\cdot;\; s)$\\
  \end{tabular}
  \vspace{1.4mm}

  \begin{tabular}{c}
    $\uparrow$
  \end{tabular}

  \begin{tabular}{|c|}
  \hline
    Covariant receptive fields with kernels
    $T(\cdot;\; s)$\\
  \hline
  \end{tabular}

  \begin{tabular}{c}
    $\mid$
  \end{tabular}

  \vspace{1.4mm}
  \begin{tabular}{c}
    Image data $f$\\
  \end{tabular}

\end{center}
\caption{Schematic overview of how the covariance properties of the
  receptive fields in the proposed receptive field model lead to covariant image
  measurements, from which truly invariant image representations can
  then be obtained by complementary selection mechanisms that
  operate over the parameters $s$ of the receptive fields
  corresponding to variations over scale, affine
  image deformations and Galilean motions.
  For pure scaling transformations, the parameter $s$ of
 the receptive fields will be a scalar scale parameter, whereas a
 covariance matrix $\Sigma$ is needed to capture more general affine image
  deformations. For spatio-temporal image data, an additional temporal scale
  parameter $\tau$ and an additional image velocity parameter $v$ are furthermore needed.}
\end{figure}

We have also described how invariance to local multiplicative
illumination transformations and exposure control mechanisms will be
automatically obtained for receptive fields in terms of spatial 
or spatio-temporal derivatives.
If we can assume that the illumination varies slowly and can be
regarded as constant over the support region of the receptive field,
the receptive field response will therefore have a direct physical
interpretation as corresponding to variations in the surface
structures of objects in environment.
Thus, the receptive field responses reflect important {\em physical
properties\/} of objects and events in the environment to support visual recognition.

It should be emphasized, however, that the model has not been
constructed to mimic mammalian vision or the vision system in other species.
Instead it is intended as an {\em idealized\/} theoretical and computational
model to capture inherent properties of basic image transformations
that any computational vision model needs to be confronted with.

Concerning limitations of the proposed approach, it should be stressed
that a basic requirement for obtaining true invariance with respect to the image
transformations according to the proposed invariance mechanisms,
is that the vision system has a sufficient number of receptive fields
over a {\em sufficient range\/} of filter parameters to support
invariance over a corresponding range of parameter variations.
Notably, such a limitation is consistent with the findings from
biological vision that 
the scale invariant properties of neurons may only hold over
finite ranges of scale variations \citep{ItoTamFujTan95-JNeuroPhys}.


It should also be noted that the invariance and covariance properties
are only guaranteed to hold if the same local approximation of the
image transformation is valid within the entire support region of the receptive field. 
Thus, complementary mechanisms can be needed to handle,
{\em e.g.\/}, discontinuities in depth, 
discontinuities in the illumination field
or specularities.

An interesting observation that can be made from the similarities 
between the receptive field families that have
been derived by necessity from the
assumptions and the receptive profiles found by cell recordings in
biological vision, is that receptive fields in the
retina, LGN and V1 of higher mammals are very close to {\em ideal\/} in 
view of the stated structural requirements/symmetry properties.
In this sense, biological vision can be seen as having adapted very
well to the transformation properties of the outside world and the
transformations that occur when a three-dimensional world is projected
to a two-dimensional image domain and being exposed to illumination variations.

Thus, image measurements in terms of receptive fields
according to the proposed model can (i)~be interpreted as 
corresponding to image features that are either
invariant or covariant with respect to basic geometric transformations
and illumination variations and can (ii)~serve as a
foundation for achieving invariant recognition of visual objects at
the system level
under variations in viewpoint, retinal size, object motion and
illumination.

From a background of the presented theory, we can therefore interpret the
receptive fields in V1 as highly dedicated computational units that are very well
adapted to enable the computation of invariant image representations
at higher levels in the visual hierarchy.

\section{Discussion}

In his recent overview of Bayesian approaches to understanding the
brain, \cite{Fri11-NeurImg} writes that ''\dots we are trying to infer the
causes of our sensations based on a generative model of the world''
and '¨\dots if the brain is making inferences about the causes of its
sensations then it must have a model of the causal relationships
(connections) among (hidden) states of the world that cause sensory
input. It follows that neuronal connections encode (model) causal
connections that conspire to produce sensory information''. He
furthermore states that an underlying message in several lines of
brain research is that the brain is regarded as ''optimal in some sense''. 

The presented theory can be seen as describing consequences of a
similar way of reasoning regarding the development of receptive fields
in the earliest stages of visual processing. 
If the brain is to handle the large natural variability in image data under basic
image transformations, such as scaling variations, viewing variations,
relative motion or illumination variations, then an optimal strategy
may be to adapt to these variabilities by making it possible to
respond to image transformations in terms of invariance or covariance
properties. 
If the receptive fields would not be covariant under basic image
transformations, then that would imply that some of the variabilities
in the information could not be appropriately captured by the vision
system, which would affect its performance. 
By in addition developing invariance properties at higher levels in a
visual hierarchy, the brain will be able to deal with natural image
transformations in a robust and efficient manner. 

Thus, the proposed theory of receptive fields can be seen as
describing basic physical constraints under which a learning based
method for the development of receptive fields will operate and
the solutions to which an optimal adaptive system may converge
to. \cite{Fie87-JOSA} as well as \cite{DoiLew05-JapCogSci} have
described how ”natural images are not random, instead they exhibit
statistical regularities” and have used such statistical regularities for
constraining the properties of receptive fields. 
Receptive field profiles have been derived by statistical methods such as
principal component analysis \citep{OlsFie96-Nature,RaoBal98-CompNeurSyst}, 
independent component analysis \citep{SimOls01-AnnRevNeurSci,HyvHurHoy09-NatImgStat} and
sparse coding principles \citep{LoePalSzi12-PLOS-CB}. 
The theory presented in
this paper can be seen as a theory at a higher level of abstraction,
in terms of basic principles that reflect properties of the environment
that in turn determine properties of the image data, without need for
explicitly constructing specific statistical models for the image
statistics. Specifically, the proposed theory can be used for
explaining why the above mentioned statistical models lead to
qualitatively similar types of receptive fields as the idealized
receptive fields obtained from our theory.

Concerning the closely related issue of how receptive fields are
distributed over the visual cortex,
\cite{KasSchLowCopWhiWol10-Science} 
have found that pinwheel density as defined from singularities in the
orientation fields of orientation hypercolumns is similar between 
species that separated evolutionary more than 65 million years ago.
By studying structural properties of self organizing systems for 
idealized neural interaction models, they showed that an overall suppressive nature 
of non-local long-range interactions is essential for the development 
of the pinwheel layout observed in carnivores and primates.
Thus, the distribution of orientation hypercolumns in the visual
cortex can be predicted from internal structural properties of 
self-organizing neural networks.
This paper presents a corresponding theoretical study of 
how the shapes of receptive field profiles found in the retina, LGN 
and the striate cortex can be predicted from structural
properties of the environment and of how invariance properties can be achieved
with a complementary assumption concerning the architecture of
complementary selection mechanisms that operate over ensembles of receptive fields.

In terms of computational modelling of vision, the proposed model for
covariant receptive fields leading to true invariance properties should 
require a significantly lower amount of training data compared to
approaches that involve explicit learning of receptive fields or
compared to computational models that are not based on explicit
invariance properties in relation to the image measurements. 
Specifically, we propose that if the aim is to build a computational
vision system that solves specific visual tasks, then a neuro-inspired
artificial vision system based on these types of provable invariance
properties should allow for more robust handling of natural imaging
variations.

\section*{Acknowledgements}

I would like to thank Benjamin Auffarth for valuable 
discussions and suggestions concerning the presentation.

The support from the Swedish Research Council, 
Vetenskapsr{\aa}det (contract 2010-4766)
and from the Royal Swedish Academy of Sciences 
as well as the Knut and Alice Wallenberg Foundation 
 is gratefully acknowledged.

\paragraph{No conflict of interest.}

This research has been conducted in the absence of any commercial 
or financial relationships that could be construed as a potential conflict of interest.

\bibliographystyle{plainnat}

{\footnotesize
\bibliography{bib/defs,bib/tlmac}
}

\end{document}